%Paper: hep-ph/9310379
%From: jenkins@su2xu1.UCSD.EDU (Elizabeth Jenkins)
%Date: Sun, 31 Oct 93 16:20:45 -0800

%%%%%%%%%%%%%%%%%%%%%%%%%%%%%%%%%%%%%%%%%%%%%%%%%%%%%%%%%%%%%%%%%%%
%                       INSTRUCTIONS
%
% This paper uses the harvmac macros. 9 files with postscript figures
% have been included as a uuencoded tar file with instructions for
% unpacking. If you have  epsf.tex, uncomment the following line
% and the postscript figures
%
%\input epsf
%
% will be included in the paper by the dvips program. If you do not
% have epsf.tex, you can print the figures out separately.
%
%%%%%%%%%%%%%%%%%%%%%%%%%%%%%%%%%%%%%%%%%%%%%%%%%%%%%%%%%%%%%%%%%%%%%%%%

\ifx\epsffile\undefined\message{(FIGURES WILL BE IGNORED)}
\def\insertfig#1{}% null macro
\else\message{(FIGURES WILL BE INCLUDED)}
\def\insertfig#1{{{
\midinsert\centerline{\epsfxsize=\hsize
\epsffile{#1}}\bigskip\bigskip\bigskip\bigskip\endinsert}}}
\fi

\input harvmac
%%%%%%%%%%%%%%%%%%%%%%%%%%%%%%%%%%%%%%%%%%%%%%%%%%%%%%%%%%%%%%%%%%%%%%
%
%  UCSD macros to overwrite some of the definitions in harvmac.tex
%  (include after harvmac.tex)
%  last modified 4/92
%
%%%%%%%%%%%%%%%%%%%%%%%%%%%%%%%%%%%%%%%%%%%%%%%%%%%%%%%%%%%%%%%%%%%%%%%
%
% modify the output routine for the little format
%
\ifx\answ\bigans
\else
\output={
  \almostshipout{\leftline{\vbox{\pagebody\makefootline}}}\advancepageno
}
\fi
%
%
% address
%

%
% grant numbers
%

%
% preprint number
%
\def\UCSD#1#2{\noindent#1\hfill #2%
\bigskip\supereject\global\hsize=\hsbody%
\footline={\hss\tenrm\folio\hss}}% restores pagenumbers
%
% abstract
%
\def\abstract#1{\centerline{\bf Abstract}\nobreak\medskip\nobreak\par #1}
%
%
% titlefont
%
%
\edef\tfontsize{ scaled\magstep3}
 \tfontsize  \tfontsize
 \tfontsize \font\titlei=cmmi10 \tfontsize
\font\titleis=cmmi7 \tfontsize \font\titleiss=cmmi5 \tfontsize
\font\titlesy=cmsy10 \tfontsize \font\titlesys=cmsy7 \tfontsize
\font\titlesyss=cmsy5 \tfontsize  \tfontsize
\skewchar\titlei='177 \skewchar\titleis='177 \skewchar\titleiss='177
\skewchar\titlesy='60 \skewchar\titlesys='60 \skewchar\titlesyss='60
%
%\def\titlefont{\def\rm{\fam0\titlerm}% switch to title font
%\textfont0=\titlerm \scriptfont0=\titlerms \scriptscriptfont0=\titlermss
%\textfont1=\titlei \scriptfont1=\titleis \scriptscriptfont1=\titleiss
%\textfont2=\titlesy \scriptfont2=\titlesys \scriptscriptfont2=\titlesyss
%\textfont\itfam=\titleit \def\it{\fam\itfam\titleit}\rm}
%
%
% math symbols
%
%---------------------------------------------------------------------
%
\def\inv{^{\raise.15ex\hbox{${\scriptscriptstyle -}$}\kern-.05em 1}}
  %prime
\def\lbar{{\lower.35ex\hbox{$\mathchar'26$}\mkern-10mu\lambda}} %lambda bar

%
%
% various slashed symbols
%
%
 % slashes a character
\def\dsl{\,\raise.15ex\hbox{/}\mkern-13.5mu D} %this one can be subscripted
\def\delsl{\raise.15ex\hbox{/}\kern-.57em\partial}
\def\Ksl{\hbox{/\kern-.6000em\rm K}}
\def\Asl{\hbox{/\kern-.6500em \rm A}}
\def\Dsl{\hbox{/\kern-.6000em\rm D}} %roman D
\def\Qsl{\hbox{/\kern-.6000em\rm Q}}
\def\gradsl{\hbox{/\kern-.6500em$\nabla$}}
%
% space and backspace in l mode
%
\def\lspace{\ifx\answ\bigans{}\else\qquad\fi}
\def\lbspace{\ifx\answ\bigans{}\else\hskip-.2in\fi} % $$\lbspace...$$
%
%     boxes an equation
%
\def\boxeqn#1{\vcenter{\vbox{\hrule\hbox{\vrule\kern3pt\vbox{\kern3pt
        \hbox{${\displaystyle #1}$}\kern3pt}\kern3pt\vrule}\hrule}}}
%
%     draw a little box (end of proof symbol)
%     e.g. \mbox{.1}{.1}
%
\def\mbox#1#2{\vcenter{\hrule \hbox{\vrule height#2in
\kern#1in \vrule} \hrule}}
%
%
%
%     curly letters
%
   %curly letters
\def\CA{{\cal A}} \def\CB{{\cal B}} \def\CC{{\cal C}} 
  \def\CG{{\cal G}} \def\CH{{\cal H}}
   \def\CL{{\cal L}}
\def\CM{{\cal M}} \def\CN{{\cal N}} \def\CO{{\cal O}} 
   \def\CT{{\cal T}}

%
%
%
%     derivatives
%
%

%

\def\bar#1{\overline{#1}}

\def\bra#1{\left\langle #1\right|}
\def\ket#1{\left| #1\right\rangle}
\def\abs#1{\left| #1\right|}

\def\darr#1{\raise1.5ex\hbox{$\leftrightarrow$}\mkern-16.5mu #1}

%
 %pound sterling
%
\def\half{{\textstyle{1\over2}}} %puts a small half in a displayed eqn
\def\frac#1#2{{\textstyle{#1\over #2}}} %puts a small fraction
%in a displayed eqn
%
%
%     various math operators
%
%

\def\Tr{\mathop{\rm Tr}}

%
%
%
%

%
%       relations
%
\def\ltap{\ \raise.3ex\hbox{$<$\kern-.75em\lower1ex\hbox{$\sim$}}\ }
\def\gtap{\ \raise.3ex\hbox{$>$\kern-.75em\lower1ex\hbox{$\sim$}}\ }
\def\gl{\ \raise.5ex\hbox{$>$}\kern-.8em\lower.5ex\hbox{$<$}\ }
\def\roughly#1{\raise.3ex\hbox{$#1$\kern-.75em\lower1ex\hbox{$\sim$}}}
%
%
%       This defines et al., i.e., e.g., cf., etc.
\def\ie{\hbox{\it i.e.}}        \def\etc{\hbox{\it etc.}}

\def\np#1#2#3{{Nucl. Phys. } B{#1} (#2) #3}
\def\pl#1#2#3{{Phys. Lett. } B{#1} (#2) #3}
\def\prl#1#2#3{{Phys. Rev. Lett. } {#1} (#2) #3}
\def\physrev#1#2#3{{Phys. Rev. } {#1} (#2) #3}
\def\ap#1#2#3{{Ann. Phys. } {#1} (#2) #3}

\relax

\def\nc{$N_c$}
\def\dim{{\,\rm dim\,}}
\def\nclimit{$N_c\rightarrow\infty$}

\def\clebsch#1#2#3#4#5#6{\left(\matrix{#1&#3\cr#2&#4\cr}\right.\left|
\matrix{#5\cr#6\cr}\right)}

\def\sixj#1#2#3#4#5#6{\left\{\matrix{#1&#2&#3\cr#4&#5&#6\cr}\right\}}
\def\bigclebsch#1#2#3#4#5#6{\left(\matrix{#1&#3\cr\noalign{\smallskip}
#2&#4\cr}\right.\left|
\matrix{#5\cr\noalign{\smallskip}#6\cr}\right)}
\def\bigsixj#1#2#3#4#5#6{\left\{\matrix{#1&#2&#3\cr
\noalign{\smallskip}#4&#5&#6\cr}\right\}}
\def\ninej#1#2#3#4#5#6#7#8#9{\left\{\matrix{#1&#2&#3\cr#4&#5&#6\cr
#7&#8&#9\cr}\right\}}
\def\Abel{\CA}
\def\semidirect{\mathbin{\hbox{\hskip2pt\vrule height 4.7pt depth -.3pt
width .25pt \hskip-1.7pt$\times$}}}
\def\bfrac#1#2{{#1 \over #2}}

\centerline{{\titlefont{The $1/N_c$ Expansion for Baryons}}}
\bigskip\medskip
\centerline{Roger Dashen, Elizabeth Jenkins, and Aneesh V.~Manohar}
\smallskip
\centerline{{\sl Department of Physics, University of California at San
Diego, La Jolla, CA 92093}}
\bigskip
\vfill
\abstract{A systematic expansion in $1/N_c$ is constructed for baryons
in QCD.  Predictions of the $1/N_c$ expansion at leading and subleading
order for baryon axial current coupling constant ratios such as $F/D$,
baryon masses and magnetic moments are derived.  The  baryon sector of
QCD has a light quark spin-flavor symmetry at leading order in
$1/N_c$.  The formalism of induced representations for contracted Lie
algebras is introduced to explain the consequences of this symmetry.
Relations are first derived for the simplest case of two light flavors
of quarks.  The generalization of the large $N_c$ expansion to $N_f = 3$
flavors is subtle and is treated using several complementary methods.
The $1/N_c$  expansion severely restricts the form of $SU(3)$ breaking in the
baryon sector.  Extrapolation of the large $N_c$ results to $N_c = 3$ permits
quantitative comparison with experimental data.  Deviations of the
measured quantities from exact spin-flavor symmetry predictions are
accurately described by the subleading  $1/N_c$ corrections. Implications of
the $1/N_c$ expansion for baryon chiral perturbation theory are discussed.}
\vfill
%\draftmode
\UCSD{\vbox{\hbox{UCSD/PTH 93-21}\hbox{hep-ph/9310379}}}{October 1993}

\newsec{Introduction}

The theory of the strong interactions is a strongly coupled theory at low
energies, with no small expansion parameter. The absence of a small
expansion parameter has frustrated attempts to compute low-energy
properties of hadrons directly in QCD.  't~Hooft realized that QCD has a
hidden parameter, \nc, the number of colors, and that the theory
simplifies in the \nclimit\ limit~\ref\thooft{G.~'t Hooft,
\np{72}{1974}{461}}. In the large $N_c$ limit, the meson sector of
QCD consists of a spectrum of narrow resonances, and meson-meson scattering
amplitudes are suppressed by powers of $1/\sqrt{N_c}$.  The analysis of the
baryon sector of QCD in the large $N_c$ limit is more subtle because
a baryon is a confined state of \nc\ quarks, and becomes a bound
state of an infinite number of quarks when \nclimit. Witten analyzed the
interactions of baryons with mesons in
large $N_c$~\ref\witten{E.~Witten, \np{160}{1979}{57}}, and showed that
the $N_c$ dependence of the baryon-meson amplitudes was the same as in a
semiclassical soliton model with coupling constant
$1/\sqrt{N_c}$. The Skyrme
model~\ref\skyrme{T.H.R.~Skyrme, Proc. Roy. Soc. A260 (1961) 127\semi
E.~Witten, \np{223}{1983}{433}}, in which the baryon is a
soliton of the low-energy chiral Lagrangian, is an explicit realization
of Witten's idea that the baryon is a semiclassical soliton.

Although the large $N_c$ limit of QCD was originally proposed as a
quantitative calculational method, predictions of this approach remained
largely qualitative in nature, with most results following primarily
from large $N_c$ power counting arguments.  Recent work
\ref\dmi{R.~Dashen and A.V.~Manohar, \pl{315} {1993}
{425}, 438}\ref\eji{E.~Jenkins,
\pl{315} {1993} {431}, 441, 447}\
on the low-energy pion interactions of baryons in large $N_c$ shows that
the large $N_c$ expansion of QCD makes definite quantitative predictions
for the static properties of baryons.  The $N_c \rightarrow
\infty$ predictions of QCD satisfy light quark spin-flavor symmetry
relations.  These symmetry relations are the same as those obtained in
the large $N_c$ Skyrme~\ref\ANW{G.S.~Adkins, C.R.~Nappi, and E.~Witten,
\np{228}{1983}{552}}\ref\bsw{J.~Bijnens, H.~Sonoda, and M.B.~Wise,
\pl{140} {1984} {421}}  and
non-relativistic quark models \ref\dgg{A.~De~R\'ujula, H.~Georgi, and
S.L.~Glashow, \physrev{D12} {1975} {147}}, which yield
identical group theoretic results in the large $N_c$
limit~\ref\manohar{A.V. Manohar,~\np{248}{1984}{19}}.  The leading
deviations from spin-flavor symmetry relations at $N_c \rightarrow
\infty$ are parametrized by $1/ N_c$-suppressed operators.  It is the
inclusion of $1/N_c$-suppressed effects which enables a quantitative
comparison of the predictions of the large $N_c$ expansion with the
physical situation of $N_c = 3$.  Whether the $1/N_c$ expansion proves
useful depends on the size of the $1/N_c$ corrections.  The $1/N_c$
expansion is particularly good for certain static quantities such as the
baryon-pion coupling constants and the isovector magnetic moments,
because there are no $1/N_c$ corrections in QCD to the spin-flavor
symmetry predictions~\dmi. Thus, the large $N_c$ predictions for the
ratios of the baryon-pion couplings and the isovector magnetic moments
are valid up to corrections of order $1/N_c^2$.  The phenomenological success
of light
quark spin-flavor symmetry predictions for these quantities is explained
by the $1/N_c$ expansion.

There are several assumptions implicit in the large $N_c$ approach to
QCD.  The principal assumption is that certain properties of QCD, such
as confinement and chiral symmetry breaking \ref\cw{S.~Coleman and
E.~Witten, \prl{45} {1980} {100} }, persist as $N_c$ is taken
to infinity. The confined hadronic states of the large $N_c$
theory are mesons and baryons, and the lowest-lying hadrons are the
pseudo-Goldstone bosons of spontaneous chiral symmetry breaking, and
baryons with $N_c$ quarks. The results derived in this paper do not require
chiral symmetry to be exact, and are valid even for non-zero quark masses. The
only requirement on the quark masses is that they remain finite as
$N_c\rightarrow \infty$. We will assume isospin symmetry is exact, for
simplicity. Given these
assumptions, the principal interactions of baryons at low energies are
pion interactions.  Since the baryon is a coherent state of $N_c$
quarks, and the pion couples to each of these quarks, the pion-baryon
axial vector coupling constant is of order $N_c$.  Each single pion-baryon
vertex is suppressed by one factor of $f_\pi$, which grows as
$\sqrt{N_c}$ in the large $N_c$ limit.  Thus, the pion-baryon vertex is
of order $\sqrt{N_c}$.  Baryon-pion scattering amplitudes involve two
baryon-pion vertices and therefore will grow like $N_c$.  This
large $N_c$ behavior of the baryon-pion scattering amplitude violates
unitarity unless there is a cancellation amongst diagrams with different
intermediate baryon states.  Thus, consistency of the large $N_c$ limit
results in cancellation conditions which relate different pion-baryon
coupling constants.  These constraint equations imply that the baryon
sector of QCD possesses a contracted light quark spin-flavor symmetry in
the large $N_c$ limit~\dmi\ref\gervais{J.-L.~Gervais and B.~Sakita,
\prl {52} {1984} {87}, \physrev{D30} {1984} {1795}}\ref\bardacki{K.~Bardacki,
\np{243} {1984} {197}}.  The contracted spin-flavor symmetry requires
that the baryon sector of large $N_c$ QCD contains an infinite tower of
degenerate states with $I = J = 1/2,\, 3/2,\, 5/2,\, \ldots$, and with pion
couplings in the precise ratios as those given by the large $N_c$ Skyrme
and non-relativistic quark models.

In this paper, we give a detailed analysis of the predictions of the
$1/N_c$ expansion for baryons, and we
provide details of some computations referred to in the
previous work~\dmi\eji. The previous results were derived in a
straightforward (but tedious) manner using Clebsch-Gordan coefficients.
In this
paper, they are rederived using more elegant group theoretical methods
from the
theory of induced representations. This formalism makes the comparison
of large $N_c$
QCD with the Skyrme model more transparent. The results for two light flavors
are extended to baryons containing strange quarks. Some of the results derived
for baryons with strange quarks are obtained using large $N_c$ {\sl but
without using
SU(3) symmetry}. These results hold irrespective of the mass of the $s$-quark,
and provide strong constraints on the pattern of $SU(3)$ breaking in the
baryon
sector.
The methods discussed in this work also apply to baryons containing a
single heavy quark in the $m_Q \rightarrow \infty$ limit.

The \nclimit\ constraint equations for the baryon-pion couplings are identical
to equations derived a long time ago in the study of strong-coupling models.
The logic of the $1/ N_c$ expansion presented here is quite different
from these other lines of reasoning, however.  Many of the old
derivations predate QCD or use arguments which are not justified in QCD.
The earliest work by Pauli and Dancoff~\ref\pauli{W.~Pauli and
S.~Dancoff, \physrev{62} {1942}
{85}}\ showed that a static $I=J=1/2$ nucleon strongly coupled to a
$p$-wave pion produces an infinite tower of baryon states with $I=J=1/2,\,
3/2,\, 5/2,\,
\ldots$, which is precisely the spectrum of large $N_c$ QCD. The constraint
equations on the pion-couplings found in large $N_c$ are the bootstrap
equations of
Chew~\ref\chew{G.F.~Chew, \prl{9} {1962} {233}}\ derived using Chew-Low
theory~\ref\chewlow{G.F.~Chew and F.E.~Low, \physrev{101} {1956} {1570}}. The
spectrum and properties of the strong coupling model were studied extensively
by Goebel~\ref\goebel{C.G.~Goebel, \prl{16} {1966} {1130}\semi C.G.~Goebel,
in {\it Quanta}, ed. by P.G.O.~Freund, C.J.~Goebel, and Y.~Nambu, University
of Chicago Press, (Chicago, 1970)}, and by Cook and
Sakita~\ref\cooksakita{T.~Cook and B.~Sakita, J. Math. Phys. 8 (1967)
708} and the reader is referred to these
papers for additional references to the earlier literature. The
constraint equations are also similar to some results of
Weinberg~\ref\weinbergi{S.~Weinberg, \prl{65}{1990} {1177},
\physrev{177}{1969}{2604}} derived by studying the high-energy behavior
of scattering amplitudes.

The organization of this paper is as follows.  Sect.~2 derives the contracted
$SU(4)$ spin-flavor algebra for baryons of large $N_c$ QCD using unitarity
constraints on
pion-baryon scattering. Sect.~3 explains the standard mathematical
technique of
induced representations which is used to construct irreducible representations
of the spin-flavor algebra for baryons.
Sect.~4 makes explicit the connection between the induced representations of
Sect.~3 and the standard collective coordinate quantization of Skyrmions.
Sect.~5 discusses the large $N_c$ counting rules for baryons. The
identification of the QCD baryons with particular irreducible representations
is justified in this section. Sect.~6 derives the pion-baryon couplings
in QCD,
including $1/N_c$ corrections. The results are derived for baryons
containing $u$, $d$ or $s$ quarks without assuming $SU(3)$ symmetry.
Sect.~7
derives the kaon-baryon couplings at leading order in $1/N_c$, and Sect.~8
derives the $\eta$ couplings. The meson-baryon couplings in the $SU(3)$ limit
are
discussed in Sect.~9. It is shown that the $F/D$ ratio for the axial currents
and baryon magnetic moments is 2/3, up to corrections of order $1/N_c^2$.
Sect.~10 discusses the
baryon masses, including $1/N_c$ corrections, but without assuming $SU(3)$
symmetry. The extension of the theory of induced representations to exact
$SU(3)$ flavor symmetry is
discussed in Sect.~11. The implications of the $1/N_c$ expansion for chiral
perturbation theory are presented in Sect.~12.  The conclusions are given in
Sect.~13. Since the
paper is rather long, we summarize the main results below.

\subsec{Summary of Main Results}

\item{$\bullet$} The $F/D$ ratio for the baryon axial currents is
determined to
be $2/3 + \CO\left(1/N_c^2\right)$, in good agreement with the experimental
value of $0.58\pm0.04$.

\item{$\bullet$} The $F/D$ ratio for the baryon magnetic moments is determined
to be $2/3 + \CO\left(1/N_c^2\right)$, in good agreement with the experimental
value of $0.72$. The difference between the $F/D$ ratios for the axial
currents
and magnetic moments is an indication of the size of $1/N_c^2$ corrections.

\item{$\bullet$} The ratios of all pion-baryon couplings are determined up to
corrections of order $1/N_c^2$, and the ratios of all kaon-baryon couplings are
determined to leading order. These results are independent of the mass of the
$s$-quark.

\item{$\bullet$} The $SU(3)$ breaking in the pion couplings must be linear
in strangeness at order $1/N_c$. This leads to an equal spacing rule for the
pion couplings, which agrees well with the data. The $SU(3)$ breaking in the
decuplet-octet transition axial currents is related to the $SU(3)$ breaking in
the octet axial currents.

\item{$\bullet$} The baryon mass relations
$$\eqalign{
&\Sigma^* - \Sigma = \Xi^* - \Xi\cr
&\frac 1 3 \left( \Sigma + 2 \Sigma^* \right) - \Lambda
= \frac 2 3 \left( \Delta - N \right)\cr
&\frac 3 4 \Lambda + \frac 1 4 \Sigma - \frac 1 2 \left( N + \Xi
\right) = - \frac 1 4 \left( \Omega - \Xi^* - \Sigma^* + \Delta
\right)\cr
&\frac 1 2 \left( \Sigma^* - \Delta \right) - \left( \Xi^* -
\Sigma^* \right) + \frac 1 2 \left( \Omega - \Xi^* \right)=0\cr
&\Sigma^*_Q - \Sigma_Q = \Xi^*_Q - \Xi^\prime_Q\cr
&\frac13\left(2\Sigma^*_Q +\Sigma_Q\right)-\Lambda_Q = \frac23\left(\Delta -
N\right)\cr
}$$
are valid up to corrections of order $1/N_c^2$ without assuming $SU(3)$
symmetry. Some of these
relations are also valid using broken $SU(3)$, with octet symmetry breaking.
Relations which can be proved using either large $N_c$ or $SU(3)$ work
extremely
well, because effects which violate these relations must break both symmetries.
The implications of $SU(3)$ breaking for the above relations are discussed more
quantitatively in Sect.~10.

\item{$\bullet$} The chiral loop correction to the baryon axial
currents
cancels to two orders in the $1/N_c$ expansion, and is of order $1/N_c$
instead of order $N_c$.

\item{$\bullet$} The order $N_c$ non-analytic correction to the baryon masses
is pure $SU(3)$ singlet, and the order one contribution is pure $SU(3)$ octet.
Thus
violations of the Gell-Mann--Okubo formula are at most order $1/N_c$. The
baryon masses can be strongly non-linear functions of the strange quark mass,
and still satisfy the Gell-Mann--Okubo formula.

\newsec{The Spin-Flavor Algebra for Baryons in Large $N_c$ QCD}

An effective light quark spin-flavor symmetry for baryons
emerges in the large $N_c$ limit of QCD.  In this section, the
spin-flavor algebra for baryons in the large $N_c$ limit is derived for
the case of $N_f=2$ light quark flavors.  The generalization of this
symmetry to three light flavors is subtle, and is addressed in later
sections.  For $N_f=2$ light flavors, large $N_c$ baryon representations
occur with the same spin and isospin quantum numbers as QCD $N_c=3$
baryon representations.  For $N_f=3$ light flavors, the $SU(3)$ flavor
representations of large $N_c$ baryons are different from $N_c=3$
baryons.  Hence, the identification of baryon states in the large $N_c$
limit with the physical baryon states of $N_c=3$ QCD is not unique.
This ambiguity complicates the discussion for $N_f=3$.  The derivation
of the contracted spin-flavor symmetry for baryons of this section
assumes that
the baryon mass is of order $N_c$, the axial vector coupling constant $g_A$ is
of order $N_c$, and the pion decay constant $f_\pi$ is of order $\sqrt{N_c}$
in the large $N_c$ limit. These assumptions are justified in Sect.~5.

The spin-flavor algebra of large $N_c$ baryons is derived by studying
the interactions of baryons with low-energy pions.  In the large $N_c$
limit, the baryon mass is order $N_c$, so a baryon is infinitely heavy
compared to a pion.  Thus, the baryon can be treated as a static
fermion, and the pion-baryon coupling can be analyzed in the rest frame
of the baryon.  Since pions are pseudo-Goldstone bosons of chiral
symmetry breaking, they are derivatively coupled to the axial vector
baryon current.  A general pion-baryon coupling is written as the baryon
axial current matrix element
\eqn\IIIi{
\bra{B^\prime} \bar q \gamma^i\gamma_5 \tau^a q \ket{B} = N_c \, g
\left(X^{ia} \right)_{B^\prime B},
}
times the derivatively coupled pion field $\partial^i \pi^a/ f_\pi$,
where the index $a$ represents the flavor index of the pion and the
index $i$ is a spin index.  The operator $X^{ia}$ has an expansion in powers of
$1/N_c$, $X^{ia}=X_0^{ia}+ X_1^{ia}/N_c + \ldots$.
The $p$-wave pion carries spin one and
isospin one.  The labels $B$ and $B^\prime$ include the spin and isospin
quantum numbers of the baryons.
The axial vector matrix element \IIIi\ is non-vanishing only for baryon
states $B$ and $B^\prime$ with spin and isospin quantum numbers combined
to form a spin one and isospin one axial vector current.  In addition,
the baryon states must be degenerate in the large $N_c$ limit, since a
soft pion will not access baryon states separated by a mass gap.  In
eq.~\IIIi, the coupling constant $g$ is chosen so that the matrix
$X_0^{ia}$ has a convenient normalization, which will be chosen later.
An explicit factor of $N_c$
is factored out of the matrix element to keep all $N_c$-dependence
manifest.  Since $f_\pi \sim \sqrt{N_c}$ in the large $N_c$ limit, the
baryon-pion vertex grows as $\sqrt{N_c}$.  The growth of the baryon-pion
vertex as $\sqrt{N_c}$ in the large $N_c$ limit results in consistency
conditions for the baryon-pion matrix elements.

Consistency conditions for pion-baryon coupling constants can be derived
by looking at the large $N_c$ behavior of pion-baryon scattering.
Consider the
scattering amplitude for the process $\pi^a(\omega,{\bf k}) + B \rightarrow
\pi^b(\omega,{\bf k^{\prime} })+ B^\prime$ in the \nclimit\ limit at
fixed pion
energy $\omega$.  The dominant diagrams in the large $N_c$ limit are shown in
\fig\fIIIi{Graphs contributing to pion-baryon scattering at leading
order in the $1/N_c$ expansion.}.  The scattering amplitude is given by
\eqn\IIIii{
\CA = -i\  {N_c^2 g^2 \over f_\pi^2}\  {k^i k^{\prime j}\over \omega}
\ \left[X^{jb}_0, X^{ia}_0\right]_{B^\prime B},
}
where the matrix product of the $X_0$'s sums over all possible baryon
intermediate states.  The commutator $\left[X^{jb}_0, X^{ia}_0\right]$
arises from a relative minus sign between the two graphs in \fIIIi\
because the intermediate baryon in \fIIIi(a) is off-shell by an energy
$\omega$, whereas the intermediate baryon in \fIIIi(b) is off-shell by
an energy $-\omega$.  The incident and emitted pion have the same
energy, since no energy can be transferred to an infinitely heavy
baryon.  Only intermediate states which are degenerate with the initial and
final baryons in the \nclimit\ limit contribute to the amplitude.  The
pion-baryon
scattering amplitude, which is in a
single partial wave, grows as $N_c$ in violation of unitarity, unless
the pion-baryon couplings satisfy the consistency
condition~\dmi\gervais\chew
\eqn\IIIiii{
\left[X^{jb}_0, X^{ia}_0\right]=0.
}

Since $X_0^{ia}$ is an irreducible tensor operator with spin one and
isospin one, it satisfies the commutation relations
\eqn\IIIiv{
\left[J^i, X^{jb}_0\right]=i\,\epsilon_{ijk}\, X^{kb}_0,\qquad
\left[I^a, X^{jb}_0\right]=i\,\epsilon_{abc}\, X^{jc}_0,
}
where $J^i$ and $I^a$ are generators of spin and isospin
transformations, respectively.  These generators satisfy the usual
commutation relations for spin and isospin,
\eqn\IIIv{
\left[J^i,J^j\right]=i\,\epsilon_{ijk}\, J^k,\qquad
\left[I^a,I^b\right]=i\,\epsilon_{abc}\, I^c,\qquad
\left[I^a,J^i\right]=0.
}
Eqs.~\IIIiii, \IIIiv\ and~\IIIv\ are the commutation relations of a
contracted $SU(4)$ algebra~\dmi\gervais\bardacki\goebel.  Consider the
embedding of the spin
$\otimes$ flavor group $SU(2)\otimes SU(2)$ in a larger $SU(4)$ group
such that the defining representation ${\bf 4}\rightarrow ({\bf 2},{\bf 2})$
under the decomposition $SU(4)\rightarrow SU(2)\otimes
SU(2)$.\footnote{${}^\dagger$}{$SU(2)$ representations will be labeled either
by
their $J$ value, or by their dimension ${\bf 2J+1}$ in boldface.} If the
generators of $SU(2)\otimes SU(2)$ in the defining representation are
$J^i$ and $I^a$, the $SU(4)$ generators in the defining representation
are proportional to $J^i\otimes 1$, $1\otimes I^a$ and $J^i\otimes I^a$,
which will be denoted by $I^a$, $J^i$ and $G^{ia}$, respectively. (The
properly normalized $SU(4)$ generators are $I^a/\sqrt{2}$,
$J^i/\sqrt{2}$ and $\sqrt2\, G^{ia}$.)  The commutation relations of
$SU(4)$ are
\eqn\IIIvi{
\eqalign{
&\left[J^i,J^j\right]=i\,\epsilon_{ijk}\,J^k,\cr
&\left[I^a,G^{jb}\right] = i\,\epsilon_{abc}\, G^{jc},\cr
&\left[I^a,J^i\right]=0,}
\qquad\eqalign{
&\left[I^a,I^b\right]=i\,\epsilon_{abc}\, I^c,\cr
&\left[J^i,G^{jb}\right] = i\,\epsilon_{ijk}\, G^{kb},\cr
&\left[G^{ia},G^{jb}\right] = \frac i 4\, \epsilon_{ijk}\ \delta_{ab}\, J^k +
\frac i 4\,\epsilon_{abc}\ \delta_{ij}\, I^c.
}}
The large $N_c$ spin-flavor algebra for baryons is obtained by taking
the limit
\eqn\IIIvii{
X^{ia}_0 = \lim_{N_c\rightarrow\infty} {G^{ia} \over N_c}.
}
This limiting procedure is known as a contraction. The only $SU(4)$
commutation relation which is affected by the contraction is
$\left[G^{ia}, G^{jb}\right]$ which is not homogeneous in $G$, and turns
into the consistency condition for pion-baryon scattering eq.~\IIIiii.
The other commutation relations become
eqs.~\IIIiv\ and \IIIv.

A better understanding of this contracted spin-flavor Lie algebra is
obtained by considering other examples of group contractions.  A simple
example of group contractions is provided by the rotation group with
commutation relations
\eqn\IIIviii{
\left[J_1,J_2\right]=i\, J_3,\quad \left[J_2,J_3\right]=i\, J_1,\quad
\left[J_3,J_1\right]=i\, J_2.
}
One possible contraction is to define $X_i=J_i/\lambda$, and take the limit
$\lambda\rightarrow\infty$. This leads to the trivial Abelian algebra
$\left[X_i,X_j\right]=0$. A more interesting contraction is to define
$P_1=J_1/\lambda$, $P_2=J_2/\lambda$, and take the limit $\lambda\rightarrow
\infty$, without rescaling $J_3$. This contraction gives the algebra
\eqn\IIIvix{
\left[P_1,P_2\right]=0,\quad \left[J_3,P_2\right]=-i\, P_1,\quad
\left[J_3,P_1\right]=i\, P_2,
}
which is the Lie algebra of motions in the $x_1-x_2$ plane where $J_3$
generates rotations about the $x_3$ axis, and $P_1$ and $P_2$ generate
translations along the $x_1$ and $x_2$ axes.  This algebra is obtained
from the rotation group which describes the motion of a point on the
two-sphere by considering a neighborhood of the
north pole of the sphere, of size $1/\lambda$, \ie\ points
$(x_1,x_2,x_3)$ where $x_3\approx1$ and $x_1$ and $x_2$ are of order
$1/\lambda$. Look at this region under a magnifying glass, so that it is
enlarged back to finite size. This magnification is equivalent to using
coordinates near the north pole of the form
$(x_1,x_2,x_3)=(y_1/\lambda,y_2/\lambda,1)$, and labeling the points by
$(y_1,y_2)$. The generators of motions on the sphere in the new
coordinates are now $J_3$ and $J_1/\lambda$ and $J_2/\lambda$.  In the
limit that $\lambda\rightarrow\infty$, the neighborhood of the north
pole becomes flat. The generator $J_3$ in this limit still generates
rotations about the $x_3$ axis, and the generators $J_1$ and $J_2$
generate translations in $y_1$ and $y_2$.

The physical interpretation of the contracted spin-flavor algebra for
baryons can be made in analogy to the above example.  In the large $N_c$
limit, the axial vector matrix elements of the baryon fields become
classical objects since the contributions of quarks in the baryon add
coherently, and the axial currents grow with $N_c$.
The operators $X_0^{ia}$ which are the rescaled axial
currents have
a well-defined large $N_c$ limit in which they become classical
commuting variables.
The normal spin and isospin symmetries of the baryon states
are thus extended to include the spin-flavor generators $X_0^{ia}$. This
extension is
possible because the baryon field is static in the large $N_c$ limit.

The large $N_c$ consistency condition for the pion-baryon couplings
eq.~\IIIiii\ determines the matrix elements up to an overall
scale. In ref.~\dmi, this solution was found by writing
$X^{ia}_0$ in terms of reduced matrix elements times Clebsch-Gordan
coefficients, and then solving the consistency conditions for the reduced
matrix elements. The same results can also be obtained by classifying all the
irreducible representations of the contracted $SU(4)$ algebra. This method
is pursued in the next section using the theory of induced
representations. The construction of induced representations is closely
related
to the quantization of Skyrmions.  The connection between large
$N_c$ QCD and the Skyrme model is discussed in Sect.~4.

\newsec{Induced Representations}

The theory of induced group representations gives a complete
classification of all irreducible representations of a semidirect
product $\CG\semidirect \Abel$ of a compact Lie group $\CG$ and an
Abelian group $\Abel$.  The contracted $SU(4)$ spin-flavor algebra for
baryons in large $N_c$ QCD is the semidirect product of an $SU(2)\otimes
SU(2)$ Lie algebra generated by ${J^i}$ and ${I^a}$, and an Abelian
algebra generated by $X_0^{ia}$. The irreducible representations of the
contracted spin-flavor algebra contain the baryon representations of
large $N_c$ QCD.  In this section, all possible irreducible
representations of the contracted spin-flavor algebra are constructed,
and the representations which describe large $N_c$ baryons are
identified. Much of the discussion is well
known~\cooksakita\ref\mackey{G.W.~Mackey, {\it Induced Representations of
Groups and Quantum Mechanics}, Benjamin, (New York, 1968)},
but may not be familiar to most physicists.

The irreducible representations of a semidirect product $\CG\semidirect
\Abel$ are induced by the irreducible representations of the Abelian
group $\Abel$.  For the contracted spin-flavor algebra of large $N_c$
baryons, the Abelian group consists of the generators $X_0^{ia}$, which
satisfy the large $N_c$ consistency condition eq.~\IIIiii.  Because the
$X_0^{ia}$ commute, states can be chosen in which $X_0^{ia}$ are treated
as coordinates,
\eqn\IVi{
\left(X^{ia}_0\right)_{\rm op} \ket{X^{ia}_0,\ldots} = X^{ia}_0
\ket{X^{ia}_0,\ldots},
}
so that the state $\ket{X^{ia}_0,\ldots}$ is labeled by the $c$-number
eigenvalues $X^{ia}_0$ of the operator $\left(X^{ia}_0\right)_{\rm op}$.
The ellipsis in the state vector represents other quantum numbers which
will be necessary to completely specify the state.  Since the states
have been chosen to diagonalize the generator
$\left(X^{ia}_0\right)_{\rm op}$, the distinction between the operator
and its eigenvalue $X_0^{ia}$ can be dropped.  To further simplify the
notation, the numbers $X_0^{ia}$ will be treated as a $3\times3$ matrix
$X_0$, whose rows are labeled by the spin index $i$ and whose columns
are labeled by the isospin index $a$.

The generators of the Abelian group $\Abel$ do not commute with the
generators of $\CG$.  The commutation relations
\eqn\IVii{
\left[J^i,X_0^{jb}\right]=i\,\epsilon_{ijk}\ X_0^{kb},\qquad
\left[I^a,X_0^{jb}\right]=i\,\epsilon_{abc}\ X_0^{jc},
}
imply that $X_0^{ia}$ is an irreducible tensor which transforms as
$(1,1)$ under the $SU(2)_{\rm spin}\otimes SU(2)_{\rm flavor}$ group.
For a finite spin
$\otimes$ flavor group transformation $(g,h)$, $X_0^{ia}$ transforms as
\eqn\IViii{\eqalign{
U_J(g)^\dagger\ X_0^{ia}\ U_J(g) &= D_{ij}^{(1)}(g)\ X_0^{ja} ,\cr
U_I(h)^\dagger X_0^{ia}\ U_I(h) &=  D_{ab}^{(1)}(h)\ X_0^{ib},
}}
where $U_J(g)$ is the unitary operator corresponding to a finite spin
rotation by the group element $g$, and $D_{ij}^{(1)}(g)$ is the rotation
matrix in the spin-one irreducible representation, since $X_0$ is an
irreducible tensor operator with spin one. The isospin transformation is
defined analogously. The spin one rotation matrix $D_{ij}^{(1)}$ is the
familiar rotation matrix for vectors in 3-dimensional space, and will be
denoted by $R$.
Eqs.~\IVi\ and \IViii\ give the action of finite rotations in
spin and isospin on the basis states $\ket{X_0,\ldots}$,
\eqn\IViv{\eqalign{
U_J(g) \ket{X_0,\ldots} &= \ket{R(g) X_0 ,\ldots^\prime},\cr
U_I(h) \ket{X_0,\ldots} &= \ket{X_0 R^{-1}(h),\ldots^\prime},
}}
where the matrix notation for $X_0^{ia}$ is used. The prime on the
unspecified labels $\ldots$ is a reminder that the spin and isospin
transformations
may affect these indices.  The infinitesimal form of
these relations shows that ${\bf J}$ and ${\bf I}$ can be represented in the $
\ket{X_0,\ldots}$ basis as differential operators
\eqn\IVv{\eqalign{
J^i &= -i\ \epsilon_{ijk}\ X_0^{jc}{ \partial\over\partial
X_0^{kc}}+\ldots,\cr
I^a &= -i\ \epsilon_{abc}\ X_0^{kb}{ \partial\over\partial
X_0^{kc}}+\ldots,\cr
}}
where the ellipsis denotes operators acting on the $\ldots$ part of
$\ket{X_0,\ldots}$.

An irreducible representation of $\CG \semidirect \Abel$ can now be
constructed.
First pick a reference state $\ket{{\bar X_0},\ldots}$ in the
irreducible representation. All other states in the irreducible
representation containing $\ket{{\bar X_0},\ldots}$ are obtained from
the reference state by applying group transformations (by the definition
of an irreducible
representation).  The reference state is an eigenvector of the generators
$X_0^{ia}$, so group transformations generated by the $X_0^{ia}$ only
change the phase of $\ket{{\bar X_0},\ldots}$, and do not produce
additional states.  Group transformations generated by ${\bf J}$ and
${\bf I}$ change the value of $X_0$, and produce new states.  Let the
state $\ket{{\bar X_0},\ldots}$
represent a point ${\bar X_0}$ in the 9-dimensional space of $3\times 3$
matrices, with coordinates $\bar X_0^{ia}$. The group transformation
$U_J(g)$ on the state $\ket{{\bar X_0},\ldots}$ gives a new point with
coordinates given by $R(g){\bar X_0}$, and similarly for $U_I(h)$. The
set of all such points as $g$ and $h$ vary over all possible $SU(2)$
matrices is called the orbit of $\bar X_0$, and is the set of all points
$R(g) {\bar X_0} R^{-1}(h)$ obtained by applying arbitrary spin and
isospin transformations $g$ and $h$, respectively, to $\bar X_0$.  Each
transformation
matrix is parameterized by three Euler angles, so the orbit of $\bar
X_0$ is at most a 6-dimensional subspace of the 9-dimensional space of
$X_0$'s.

The 9-dimensional space of all possible $3\times3$ matrices $X_0^{ia}$
can be divided up into different disjoint orbits.  Each orbit
corresponds to a different irreducible representation, since states are
in the same orbit if and only if they are connected by a group
transformation.  The different orbits can be classified in a simple
manner.  Any matrix $\bar X_0$ can be brought to real diagonal form,
\eqn\IVvi{
\bar X_0 \rightarrow
\pmatrix{\lambda_1&0&0\cr0&\lambda_2&0\cr0&0&\lambda_3\cr}
}
by a transformation $\bar X_0 \rightarrow R(g) \bar X_0 R^{-1}(h)$, with
at most one $\lambda_i$ being negative\footnote{${}^\dagger$}{The sign
of any two
eigenvalues can be changed simultaneously, but the sign of a single
eigenvalue cannot be changed since ${\rm det}\, R=1$.} and $\lambda_1\le
\lambda_2 \le \lambda_3$.  The reference point $\bar X_0$ on each orbit
can be chosen to be a matrix in the standard form eq.~\IVvi.  Any
rescaling of the $X_0$'s can be reabsorbed into a redefinition of the
coupling constant $g$ of eq.~\IIIi, so  a normalization convention for
the $X_0$'s can be imposed,
\eqn\IVvii{
X_0^{ia} X_0^{ia} = \Tr X_0^2 = 3.
}
This normalization condition restricts the $\bar X_0$'s to have
$\lambda_1^2+\lambda_2^2+\lambda_3^2=3$. Any $\bar X_0$ other than the
trivial representation with all $\lambda_i=0$ can be brought into this
standard form by a trivial rescaling.  The trivial representation which
only contains $X_0 =0$ is not of interest for the problem under study,
since it corresponds to vanishing pion-baryon couplings.  Hence, all
non-trivial irreducible representations are classified by $\bar X_0$'s
of the form eq.~\IVvi\ with $\Tr \bar X_0^2=3$.

The unspecified labels $\ldots$ of the irreducible representation
transform under the little group of $X_0$.  The little group $\CG_X$ of
a point $X_0$ on an orbit is the set of $SU(2)_{\rm spin} \otimes
SU(2)_{\rm isospin}$ transformations $(g,h)$ which leave the point $X_0$
unchanged, so that $R(g) X_0 R^{-1}(h)=X_0$.  Since the different points
on an orbit are connected to each other by group transformations, the
little groups at different points on the same orbit are isomorphic.  For
example, suppose the point $X_0^\prime$ is obtained from $X_0$ by the
group transformation $(g_0,h_0)$, \ie\ $X_0^\prime = R(g_0) X_0
R^{-1}(h_0)$.  The transformation $(g^\prime,h^\prime)$ leaves
$X_0^\prime$ invariant if and only if $(g,h)$ leaves $X_0$ invariant,
where $g^\prime=g g_0$ and $h^\prime=h h_0$.
Any point $X_0$ on the orbit of
$\bar X_0$ can be obtained from $\bar X_0$ by a group transformation
$(g, h) \in \CG$.  If $(g,h)$ transforms $\bar X_0$ into $X_0$, then so
does $(g,h)(g^\prime,h^\prime)_l$, where $(g^\prime,h^\prime)_l$ is an
element of the little group $\CG_{\bar X_0}$ of $\bar X_0$.  Thus
elements on the orbit of $\bar X_0$ are in one-to-one correspondence
with elements of the coset space $\CG/ \CG_{\bar X_0}$.

It is easy to work out the
little group at the standard configuration eq.~\IVvi.  If all the
$\lambda_i$ are different, the little group consists of $Z_2\times Z_2$
where the $Z_2$'s are generated by $2\pi$ rotations $U_J(2\pi)$ and
$U_I(2\pi)$ in spin and isospin, respectively.  If two of the
$\lambda_i$'s are equal and non-zero, the little group is $U(1)\times
Z_2$, where the $U(1)$ is spin-isospin rotations in the two-plane of
degenerate eigenvalues with $g=h$, and the $Z_2$ is a $2\pi$ rotation in
space. If two of the $\lambda_i$'s are equal and zero, the little group
is $U(1)\times U(1)$, with independent $U(1)$ rotations in spin and
isospin. If three of the eigenvalues are equal and non-zero, the little
group is $SU(2)\times Z_2$, where the $SU(2)$ is spin-isospin rotations
with $g=h$ and the $Z_2$ is a $2\pi$ rotation in space.  If all three
eigenvalues are zero, the little group is $SU(2)\times SU(2)$.

If the little group is non-trivial, the additional state labels $\ldots$
correspond to irreducible representations of the little group.  The
large $N_c$ baryon states which will be considered extensively in this
paper are irreducible representations which belong to the orbit with
standard configuration
\eqn\IVviii{
\bar X_0 = \pmatrix{1&0&0\cr0&1&0\cr0&0&1\cr},
}
and little group $SU(2)\times Z_2$.  The generators of the $SU(2)$
little group of $\bar X_0$ are
\eqn\IVix{
{\bf K} = {\bf I} + {\bf J}\, ,
}
and the irreducible representations are $\ket{ K, k}$ with $K = 0$,
$1/2$, $1$,
$\ldots$  The irreducible representations of $Z_2$ are $\pm$, depending on
whether the state is even or odd under $2\pi$ rotations, \ie\ whether it is
bosonic or fermionic.  The state $\ket{ {\bar X_0}, K, k, \pm}$ transforms as
an irreducible representation under the little group,
\eqn\IVx{\eqalign{
&U_K(g) \ket{ {\bar X_0}, K, k, \pm} = \ket{ {\bar X_0}, K, k, \pm}
D^{(K)}_{k^\prime k}(g), \cr
&U_J(2\pi) \ket{ {\bar X_0}, K, k, \pm} = \pm \ket{ {\bar X_0}, K, k,
\pm}, \cr
}}
where $D^{(K)}$ are the $SU(2)$ rotation matrices in the $(2K+1)$
dimensional representation.  Note that the value of $\bar X_0$ is
unchanged by the action of the little group.  The irreducible
representation of the little group eq.~\IVx\ induces an irreducible
representation of the full symmetry group when combined with the
transformation of $X_0$, eq.~\IViii.  The discrete $Z_2$ label separates the
baryon representations into fermions and bosons.
For $N_c$ odd, baryon states are fermions and the $Z_2$ label is restricted to
be $-$.  The $N_c$ even case corresponds to baryon states which are
bosons, and
$Z_2$ label $+$.  This case is not of physical interest since $N_c =3$ in QCD.
For the remainder of this paper, $N_c$ is understood to be odd, and the $Z_2$
label of the states and the $Z_2$ factor of the little group are omitted.

It is useful to specify a single group element in $\CG$
connecting each point $X_0$ on the orbit of $\bar X_0$ with the reference
point $\bar X_0$.  For the special case of interest, the group
transformations are of the form $(g,h)$, and the little group is the set
of elements $(g,g)$.  A convenient choice in this case is to use the
group element $(1,hg^{-1})$
to represent a point on the orbit with $X_0=R(g)\bar X_0 R^{-1}(h)$. (Another
possible choice is to use $(gh^{-1},1)$ as a representative element.)  The
states  $\ket{{ X_0},K,k}$ are defined by
\eqn\IVxi{
\ket{{ X_0},K,k} = U_I(hg^{-1}) \ket{{\bar X_0},K,k},
}
where $X_0=R(g) \bar X_0 R^{-1}(h)$.
The connection eq.~\IVxi\ relates the basis states $\ket{K,k}$ at the
reference point $\bar X_0$ to the basis states at point $X_0$ of the
orbit.  The connection is convention dependent, but different
conventions are equivalent.  A familiar example of such a problem is the
motion of a spin-1/2 particle on a sphere, with the spin of the particle
constrained to be tangential to the sphere. Two tangential directions,
say $\hat x$ and $\hat y$, can be chosen at the north pole of the
sphere. The tangential directions at other points are defined by
parallel transport. Different paths from the north pole to a given point
$P$ give different definitions of the tangent vectors at $P$, and a
standard path must be chosen to define basis vectors. Although the
connection is convention dependent, the tangent plane spanned by the
basis vectors at $P$ is
well-defined, and independent of the choice of path. This example is
analogous to the definition eq.~\IVxi. The state $\ket{{ X_0},K,k}$
depends on the choice of group transformation to go from $\bar X_0$ to
$X_0$, but the space of linear superpositions $\sum_k c_k \ket{{
X_0},K,k}$ is well-defined at each point $X_0$. (A more elegant
formulation of this construction in terms of fiber bundles is left to
the reader.)

Eqs.~\IVx\ and \IVxi\ determine the transformation law for
$\ket{X_0,K,k}$ under a general group transformation $(g,h)$.  Let $X_0$
be obtained from $\bar X_0$ by the transformation $(1,g_X)$.  Then
\eqn\IVxii{\eqalign{
&U_J(g) U_I(h) \ket{X_0,K,k} = U_J(g) U_I(h) U_I(g_X) \ket{{ \bar
X_0},K,k}\cr
&=U_I(h g_X g^{-1})  U_I(g) U_J(g) \ket{{\bar X_0},K,k}
=U_I(h g_X g^{-1})  U_K(g) \ket{{\bar X_0},K,k}\cr
&=U_I(h g_X g^{-1}) \ket{{\bar X_0},K,k^\prime} D_{k^\prime k}^{(K)}(g)
= \ket{\bar X_0 R^{-1}(h g_X g^{-1}),K,k^\prime} D_{k^\prime k}^{(K)}(g)\cr
&= \ket{R(g) X_0 R^{-1}(h),K,k^\prime} D_{k^\prime k}^{(K)}(g).\cr
}}
Eq.~\IVxii\ generalizes the transformation law eq.~\IViv\ to
include the transformation of the $\ket{K,k}$ part of the state.
The generalization of eq.~\IVv\ is obtained by taking the infinitesimal
form of
eq.~\IVxii,
\eqn\IVxiii{\eqalign{
J^i &= -i\ \epsilon_{ij\ell}\ \delta_{ k^\prime k}\ X_0^{jc}{ \partial
\over \partial X_0^{\ell c}} + T^{(K)i}_{k^\prime k},\cr
I^a &= -i\ \epsilon_{abc}\ \delta_{k^\prime k}\ X_0^{\ell b}{
\partial\over\partial
X_0^{\ell c}},\cr
}}
where $T^{(K)}$ are $SU(2)$ generators in the $(2K+1)$ dimensional
representation. The asymmetry between ${\bf J}$ and ${\bf I}$ in
eq.~\IVxiii\ occurs because group elements of the form $(1, gh^{-1})$
were chosen to represent the coset.  If the convention $(hg^{-1}, 1)$
had been adopted instead of $(1, gh^{-1})$, the $T^{(K)}$ matrices would
have been added to ${\bf I}$ instead of to ${\bf J}$.
The basis states can be chosen to have a group-invariant normalization.
If $(1,g)$ transforms
$\bar X_0 \rightarrow X_g$, then the states are normalized so that
\eqn\IVxiv{
\bra{X_h,K,k^\prime}\left.X_g,K,k\right\rangle = \delta_{k^\prime k}
\ \delta(gh^{-1})
}
where $\delta(g)$ is a $\delta$-function on the $SU(2)$ group normalized so
that
$\int dg \,\delta(g)=1$.

A similar construction works for the other orbits, which have different
little groups.
The above recipe for constructing the irreducible representations of the
large $N_c$
spin-flavor symmetry group can be summarized: (i) pick an orbit  (ii) find the
little group of the orbit and choose an irreducible representation of the
little group. The irreducible representation of the little group induces
an irreducible representation of the full group.  The importance of this
construction is due to a theorem of Mackey, which states that the above
procedure yields all the irreducible representations of the semidirect
product group $\CG\semidirect
\CA$~\mackey.\footnote{${}^\dagger$}{The proof of Mackey's
theorem depends on a technical assumption that one can find a
Borel set which contains exactly one point in each disjoint orbit. This
assumption is valid for the large $N_c$ spin-flavor group.}

The induced representations constructed above are the different solutions to
the consistency equations for pion-baryon scattering. The solutions are given
in terms of basis states $\ket{X_0,K,k}$ which diagonalize the axial
currents. It is more convenient to work with states of definite spin and
isospin, \ie\ in a basis in which ${\bf J}$ and ${\bf I}$ are diagonal,
since baryon states have definite spin and isospin. Let $X_g$ denote the
value of $X_0$ obtained by acting  on $\bar X_0$ with the group
transformation $(1,g)$. States of definite isospin are obtained by
taking linear superpositions of $\ket{X_g,K,k}$,
\eqn\IVxv{
\ket{I\, I_3 m ; K k} = \int\ dg\, D^{(I)}_{I_3 m}(g)^* \ket{X_g,K,k}.
}
Since the representation matrices $D^{(I)}_{I_3 m}(g)$ form a complete
set of basis functions on the group, the states $\ket{I \,I_3 m ; K k}$
are complete.
Under an isospin rotation, $\ket{I \, I_3 m ; K k}$ transforms as
\eqn\IVxvi{\eqalign{
&U_I(h) \ket{I \, I_3 m; K k} = \int\ dg \, D^{(I)}_{I_3 m }(g)^* U_I(h)
\ket{X_g,K,k}\cr
&=\int\ dg \, D^{(I)}_{I_3 m }(g)^* \ket{X_{hg},K,k}
=\int\ dg \, D^{(I)}_{I_3 m }(h^{-1}g)^* \ket{X_g,K,k}\cr
&=\int\ dg \, D^{(I)}_{I_3 I_3^\prime}(h^{-1})^* D^{(I)}_{I_3^\prime
m}(g)^* \ket{X_g,K,k}
= \ket{I \, I_3^\prime m ; K k}D_{I_3^\prime I_3}^{(I)}(h),\cr
}}
where the second line follows from eq.~\IVxii\ and the invariance of the group
measure. Eq.~\IVxvi\ implies that $\ket{I \,
I_3 m ; K k}$ is a state which transforms under isospin as $\ket{ I\,
I_3 }$. A similar calculation can be done for a spin transformation,
\eqn\IVxvii{\eqalign{
&U_J(h) \ket{I \, I_3 m ; K k} = \int\ dg \, D^{(I)}_{I_3 m }(g)^*
U_J(h) \ket{X_g,K,k}\cr
&=\int\ dg \, D^{(I)}_{I_3 m }(g)^*
\ket{X_{gh^{-1}},K,k^\prime}D^{(K)}_{k^\prime k}(h)\cr
&=\int\ dg \, D^{(I)}_{I_3 m }(gh)^* \ket{X_g,K,k^\prime}D^{(K)}_{k^\prime
k}(h)\cr
&=\int\ dg \, D^{(I)}_{I_3 m^\prime }(g)^* D^{(I)}_{m^\prime m }(h)^*
\ket{X_g,K,k^\prime}D^{(K)}_{k^\prime k}(h)\cr
&= \ket{I \, I_3 m^\prime ; K k^\prime}D_{m^\prime m }^{(I)}(h)^*
D^{(K)}_{k^\prime k}(h) \, .\cr
}}
Eq.~\IVxvii\ implies that $\ket{I \, I_3 m ; K k}$ transforms under
rotations like the product of state $\ket{Kk}$ and the complex
conjugate of the state $\ket{I m}$. States which transform
under spin rotations like $\ket{J J_3}$ are obtained by combining the $k$
and $m$ indices using Clebsch-Gordan coefficients,
\eqn\IVxviii{\eqalign{
\ket{I \, I_3, J \, J_3 ; K} &= \sqrt{{\dim I \dim J\over \dim K}}
\clebsch J {J_3} I {m} K k \ket{I \, I_3 m ; K k} \cr
\noalign{\medskip}
&=\sqrt{{\dim I \dim J\over \dim K}}
\clebsch J {J_3} I {m} K k \int\ dg \, D^{(I)}_{I_3 m}(g)^*
\ket{X_g,K,k},
}}
with an implied sum over $m$ and $k$.
The normalization factor has been chosen so that the states \IVxviii\
are normalized to unity, using the normalization eq.~\IVxiv\ for the
basis states $\ket{X_g,K,k}$.

Baryon representations can be identified with the irreducible
representations of definite spin, isospin and $K$ given in eq.~\IVxviii.
The baryon states with a given $K$ contain all states of the form
$(J,I)$, provided $J\otimes I \in K$.  The induced representations with
$K=0$ consist of an infinite
tower of states with $(J,I)=(1/2,1/2)$, $(3/2,3/2)$, $(5/2,5/2)$,
$\ldots$, $(N_c/2, N_c/2)$.  The induced representations with $K=1/2$
correspond to an infinite tower of states $(1/2,0)$, $(1/2,1)$,
$(3/2,1)$, $(3/2,2)$, $(5/2,2)$, $\ldots$; the induced representations
with $K=1$ contain the states $(1/2,1/2)$, $(3/2,1/2)$, $(1/2,3/2)$,
$(3/2,3/2)$, $(5/2,3/2)$, $(3/2,5/2)$, $(5/2,5/2)$, $\ldots$; and the
induced representations with $K=3/2$ include the states $(3/2,0)$,
$(1/2,1)$, $(3/2,1)$, $(5/2,1)$, $\ldots$. Pions
with $p$-wave couplings to baryons carry the $(J,I)$ quantum numbers
$(1,1)$ and $K=0$.  Thus, pions only connect baryons within a given $K$
sector.  The quantum numbers of the baryon states in the $K$ sectors can
be identified with the known baryon spin-$1/2$ octet and spin-$3/2$
decuplet states of QCD if the quantum number $K$ labels baryon sectors
with differing strangeness.  The $K=0$ sector contains strangeness zero
baryons such as the nucleon, which is identified with the state
$(1/2,1/2)$, and the $\Delta$, which corresponds to the state
$(3/2,3/2)$.  The $K=1/2$ sector contains the strangeness $-1$ baryons:
the $\Lambda (1/2,0)$, $\Sigma (1/2,1)$ and $\Sigma^* (3/2,1)$.  The
$K=1$ sector contains strangeness $-2$ baryons: the $\Xi (1/2,1/2)$ and
$\Xi^* (3/2,1/2)$; and the $K=3/2$ sector contains strangeness $-3$
baryons such as the $\Omega^- (3/2,0)$.  The other states in the towers
correspond to baryons which exist for $N_c \rightarrow \infty$, but not
for $N_c = 3$. The identification of baryons with
different irreducible representations of the spin-flavor group is
discussed in more detail in Sect.~5.

\newsec{Skyrmions}

The $SU(2)$ Skyrme model  provides an explicit realization of the contracted
$SU(4)$ spin-flavor algebra of Sect.~2.  The induced representations of
Sect.~3
are in one-to-one correspondence with soliton solutions of the Skyrme model.
The $SU(2)$ Skyrme model with the spherical hedgehog solution $\Sigma_0$
corresponds to the induced representation with $K=0$ and little group
$SU(2)\times Z_2$.  The $SU(2)$ Skyrmion is a soliton of the chiral Lagrangian
of the form
\eqn\Vi{
\Sigma_0({\bf x}) = e^{i\tau\cdot\hat x F(r)}\ ,
}
where $F(0)=-\pi$ and $F(\infty)=0$.\footnote{${}^\dagger$}{There is a sign
error in the choice
of $F$ in ref.~\ANW.}
The Skyrmion configuration
$\Sigma_0$ corresponds to the reference state $\ket{\bar X_0,0,0}$, which is
invariant
under ${\bf K} ={\bf I}+{\bf J}$,
\eqn\Vii{
\left({\bf I}+{\bf J} \right)\Sigma_0 =0 \ .
}
An isospin transformation of the soliton $\Sigma_0$ gives an equivalent
soliton
solution with $\Sigma=A\Sigma_0 A^{-1}$.  Spin and isospin transformations
generate
infinitesimal body and space centered rotations on the soliton $\Sigma$, with
\eqn\Viii{
A\rightarrow A U^{-1},\qquad A\rightarrow U A,
}
respectively.
The axial vector current in the Skyrme
model is equal to
\eqn\Viv{
A^{ia} = \half N_c\, g_A  \Tr \left( A \tau^i A^{-1} \tau^a \right)
}
to leading order in $N_c$, where the coupling constant $g_A$ is a function of
the shape function $F(r)$ of the soliton.
Thus, the $K=0$ soliton states $\Sigma=A\Sigma_0 A^{-1}$ correspond to the
states $\ket{X_0,0,0}$ with
\eqn\Vv{
X_0^{ia} = \half \Tr \left( A \tau^i A^{-1} \tau^a \right) .
}
The reference point of the representation, $\bar X_0^{ia} = \delta^{ia}$,
corresponds to the standard soliton configuration $\Sigma_0$ with $A=1$.
The collective coordinate $A$ of the soliton determines the coordinate
$X_0^{ia}$ by eq.~\Vv. The commutation relation \IIIiii\ is satisfied in the
Skyrme model because the collective coordinate $A$ is a $c$-number.

Spherical hedgehog solutions with $K\not=0$ correspond to the induced
representations $\ket{X_0, K, k}$.  Skyrme model solutions with $K\not=0$ are
constructed below.  The approach which is adopted is closely related to a
method introduced by Callan and Klebanov~\ref\callan{C.~Callan and
I.~Klebanov,
\np{262}{1985}{365}\semi C.~Callan, K.~Hornbostel and I.~Klebanov,
\pl{202}{1988}{260} }.  Callan and Klebanov showed that Skyrmions containing a
single strange quark can be treated as bound states of an $SU(2)$ Skyrmion and
a $K$ meson.  This method allows strange baryons to be studied in the Skyrme
model without assuming $SU(3)$ symmetry.  The Skyrme representation of baryons
containing strange quarks given below considers bound states of Skyrmions and
$s$-quarks.

First consider a baryon containing a single strange quark as a bound state of
an $SU(2)$ soliton and a $K$ meson. The $SU(2)$ soliton, which is
a fermion, can be combined with the $\bar u$ or $\bar d$ antiquark in the $K$
meson to produce a bosonic $SU(2)$ soliton, which is a color $\bar 3$.  The
strange quark baryon is the colorless bound state of this bosonic
soliton and a
strange quark.  Because the color indices of the bosonic soliton and $s$-quark
must be contracted, it is possible to instead regard the strange quark as a
colorless bosonic object with spin-$1/2$ and the bosonic soliton as a color
singlet state. Baryons with $N_s$ strange quarks arise as bound states of a
fermionic or bosonic soliton and $N_s$ strange quarks,
\eqn\Vvi{
\ket{\Sigma_0}\ket{sss\ldots s},
}
where each strange quark carries spin $1/2$. The soliton-quark bound
state is completely symmetric under the exchange of the $s$-quarks, so
$\ket{sss\ldots s}$ has the strange quark spins combined into total spin
$N_s/2$. The induced representation $\ket{X_0,K,k}$ can be identified
with a Skyrmion bound to $2K$ strange quarks,
\eqn\Vvii{
\ket{X_0,K,k}\leftrightarrow \ket{A \Sigma_0 A^{-1}}\ket{sss\ldots s},
}
where $X_0$ is related to $A$ by eq.~\Viv, and the spins of the $2K$
$s$-quarks are combined into a state with spin $K$ and spin eigenvalue
$k$. The
operators
\eqn\Vviii{\eqalign{
J^i_{ud} &= -i\,\epsilon_{ij\ell}\,\delta_{ k^\prime k}\ X_0^{jc}{ \partial
\over \partial
X_0^{\ell c}},\cr
I^a &= -i\,\epsilon_{abc}\,\delta_{k^\prime k}\ X_0^{\ell b}{
\partial\over\partial
X_0^{\ell c}},\cr
}}
generate infinitesimal space and body centered rotations of the soliton,
and can be interpreted as the spin of the light degrees of freedom ($u$
and $d$
quarks, gluons, orbital angular momentum, \etc), and the isospin,
respectively.  The operator
\eqn\Vix{
J_s^i= T^{(K)i}_{k^\prime k},
}
is the strange quark spin, and acts only on the strange quarks. The
total angular momentum $J=J_{ud}+J_s$ reproduces eq.~\IVxiii.

Quantization of non-spherical soliton solutions of the chiral
Lagrangian yields the induced representations of Sect.~3 with general reference
points eq.~\IVvi. These solutions are unimportant for the study of the
lowest-lying baryons because a non-spherical soliton differs in mass from the
spherical soliton by an amount of order $N_c$.

\newsec{Large $N_c$ Counting Rules}

The $1/N_c$ expansion for baryons relies heavily on large $N_c$ power counting
rules for baryon scattering processes and matrix elements.  In this section,
the large $N_c$ behavior of $f_\pi$, $M$ and $g_A$ are presented.  In
addition, the identification of the lowest lying baryon states with the
induced representations of Sect.~3 is discussed from a quark model approach.
Some of the results in this section are well known, and can be found in
refs.~\thooft, \witten\ and \ref\coleman{See for example, S.~Coleman, {\it
Aspects of Symmetry}, Cambridge University Press, (Cambridge, 1985)}.

\subsec{Meson Green Functions}
First consider the large $N_c$ dependence of meson Green functions.
The pion is created from the vacuum by a color singlet axial vector quark
bilinear,
\eqn\VIi{
A^{\mu a}=\sum_{\alpha=1}^{N_c} \bar q^\alpha \gamma^\mu \gamma_5 \tau^a
q_\alpha,
}
where $\tau^a$ is a flavor matrix, and the sum on colors is shown
explicitly. The two-point function $\bra{0} A^{\mu a} A^{\nu b}
\ket{0}$ is dominated in the large $N_c$ limit by planar graphs bounded by a
single quark line, as shown in \fig\fVIi{The leading diagram for the two
point function of two axial currents in large $N_c$. Only the quark line is
shown.  An arbitrary number of internal gluons can be exchanged between the
quark lines without changing the $N_c$-dependence of the diagram.},
and is order $N_c$.
The axial current two-point function is related to the pion decay constant
\eqn\VIii{
N_c \sim \int d^4x\ e^{i p \cdot x} \bra{0} A^{\mu a}(x) A^{\nu b}(0)
\ket{0} = -{ f_\pi^2  p^\mu p^\nu\over p^2} + \ldots,
}
where the ellipsis denotes terms other than the single pion pole. The
omitted terms cannot cancel the pion pole, so $f_\pi^2 \sim
N_c$. In other words, $A^{\mu a}/\sqrt{N_c}$ creates pions from the
vacuum with an amplitude that is finite as \nclimit.

The above argument
does not depend on the special form of the axial current; any two-point
function of quark bilinears is order $N_c$, so any quark bilinear
creates a meson with amplitude $\sqrt{N_c}$. This fact immediately
implies that
multi-meson vertices are suppressed by powers of $\sqrt{N_c}$. For
example, an $n$-meson amplitude is obtained by studying the $n$-point
function of quark bilinears. The dominant graphs are of the form shown
in \fig\fVIii{The dominant diagram for the $n$-point function in the large
$N_c$ limit. An arbitrary number of internal gluons can be exchanged.}, and are
proportional to $N_c$. Each quark bilinear produces a
meson with amplitude $\sqrt{N_c}$, so that the $n$-meson amplitude is of
order $N_c/(\sqrt{N_c})^n\sim N_c^{1-n/2}$. Each additional meson
produces a suppression factor of $1/\sqrt{N_c}$ in a multi-meson
amplitude.

\subsec{Baryon Green Functions}
Large $N_c$ baryons are color singlet states containing $N_c$
quarks with color indices
contracted using the $N_c$-index $\epsilon$-symbol of $SU(N_c)$. The
matrix element of a quark bilinear such as the the axial current $A^{\mu
a}$ between baryon states is given by the graphs in \fig\fVIiii{A
diagram contributing to the matrix element of the axial current between
baryon states. The axial current can be inserted on any of the $N_c$ quark
lines.}, where the
operator can be inserted on any of the $N_c$ quark lines. Each of the
insertions gives a contribution of order one, so the net contribution
from the $N_c$ diagrams is at most of order $N_c$. Note that
the contribution is {\sl at most} of order $N_c$, since there may be
cancellations amongst the $N_c$ diagrams.

We assume that the lowest lying baryon states are completely symmetric in the
spatial coordinates of the quarks, and therefore
must be completely symmetric in spin $\otimes$ flavor, since
they are totally antisymmetric in color and the quarks are fermions.
It is useful to replace the quark fields by equivalent color singlet spin-1/2
boson fields which carry the spin and flavor
indices of the original quarks. This convention is merely a notational
convenience to obtain the correct spin-flavor quantum numbers of the baryons
while avoiding color indices on the quark fields. The baryon annihilation
operator is $q_{\iota_1 \alpha_1}q_{\iota_2 \alpha_2} \ldots
q_{\iota_{N_c} \alpha_{N_c}}$ in this new notation, where $\iota_i$ and
$\alpha_i$ are the spin and flavor indices of the $i^{\rm th}$ quark,
respectively.

The $N_c$ dependence of certain baryon Green functions can be obtained
by using a trick. Assume that the baryon mass is non-zero in the large $N_c$
limit, so that one can go to the baryon rest frame. In the baryon rest
frame, one can split the quark spinor field into its upper two and lower
two components denoted  by
superscripts $(\pm)$, respectively, using the projector $(1\pm\gamma^0)/2$.
Consider
$q_{\iota \alpha}^{(+)}$ where the spinor index $\iota$ now takes on the
values $1,2$ corresponding to the two upper components of the spinor
field. (Note that there is no assumption that
the quark $q$ is non-relativistic.)
Define the baryon annihilation operator
\eqn\VIiii{
B_{\Omega}=q_{\iota_1 \alpha_1}^{(+)} q_{\iota_2 \alpha_2}^{(+)} \ldots
q_{\iota_{N_c} \alpha_{N_c}}^{(+)} \Omega_{\iota_1 \alpha_1} \Omega_{\iota_2
\alpha_2} \ldots
\Omega_{\iota_{N_c} \alpha_{N_c}},
}
where $\Omega_{\iota \alpha}$ is an arbitrary $2\times2$ matrix with one
spin index $\iota$ in the spin-1/2 representation and one flavor index
$\alpha$ in the isospin-1/2 representation. Baryons containing strange quarks
will
be discussed later in this section.
The baryon state annihilated by this operator will be denoted by $\ket{
\Omega}$. In the rest frame of the baryon, one has $SU(2)$ rotational
symmetry and $SU(2)$ flavor symmetry. (The Lorentz boost symmetry
has been broken by the choice of a particular reference frame.) $\Omega$
transforms as
\eqn\VIiv{
\Omega_{\iota_1 \alpha_1} \rightarrow D_{\iota_1 \iota_2}^{(1/2)}(g)\
D_{\alpha_1 \alpha_2}^{(1/2)}(h)\
\Omega_{\iota_2 \alpha_2},
}
under arbitrary spin
and isospin transformations $g$ and $h$, respectively.
In the large $N_c$ limit, the state $\ket{ \Omega}$ is orthogonal
to $\ket{\Omega^\prime}$ if $\Omega\not=\Omega^\prime$. The overlap
$\bra{\Omega}\left.{\Omega^\prime}\right\rangle$ involves the overlap
of the quark
$q_{\iota \alpha}\Omega_{\iota \alpha}$ with the quark
$q_{\iota \alpha}\Omega^\prime_{\iota \alpha}$ to the $N_c^{\rm th}$
power. The
quark
overlap is less than one if the two quarks are in different (not
necessarily orthogonal) states, and so the baryon overlap vanishes in
the limit \nclimit.

It is instructive to consider a simpler example which illustrates the utility
of the $\ket{\Omega}$ states.
Consider
the case of a single heavy quark flavor, so that the quark can be
treated as non-relativistic. One can then define the $N_c$-quark state
$\ket{\hat n}$ as the state annihilated by $q_{\hat n} q_{\hat n}
\ldots q_{\hat n}$, where $q_{\hat n}$ annihilates a quark with spin up
along the $\hat n$ direction. The overlap of a spin-1/2 state with spin
up along $\hat n$ and a spin-1/2 state with spin up along $\hat n ^\prime$ is
$\cos\theta/2$ (up to a phase), where $\theta$ is the
angle between $\hat n$ and $\hat n^\prime $. Thus the overlap
\eqn\VIv{
\abs{\bra{\hat n}\left.\hat n^\prime\right\rangle} =
\abs{\cos\theta/2}^{N_c}
\mathop{\longrightarrow}_{N_c\rightarrow\infty} 0\ \ \ {\rm if}\ \theta \not=
0.
}
The state $\ket{\hat z}$ is the state $\ket{\uparrow \uparrow\ldots
\uparrow}$ with $J=N_c/2$ and $J_3=N_c/2$, and is the highest weight
state. All other states with $J=N_c/2$ can be obtained from $\ket{\hat
z}$ by applying spin lowering operators. One could equally well have
started from the state $\ket{\hat n}$, which corresponds to picking the
state with the largest value of $J_3$ along the $\hat n$ axis, and
obtained all
other states by applying lowering operators along the $\hat n$ axis.
Similarly, the state $\ket{\Omega}$ defined above can be thought of as a
spin-flavor
highest weight state, from which the other spin and flavor states are
obtained by applying spin and flavor lowering operators.

The matrix element of the Hamiltonian between highest weight states
$\ket{\Omega}$ is
given by inserting the Hamiltonian on any one of the quark lines. There
are $N_c$
graphs which add constructively since all $N_c$ quarks are identical. Thus,
$\bra{\Omega} \CH \ket{\Omega}$ is of order $N_c$.
Any  operator that acts on a finite number of quarks (such as the Hamiltonian
which is a single quark operator) cannot affect the orthogonality of
$\ket{\Omega}$ for
different values of $\Omega$, so
\eqn\VIvi{
\bra{\Omega^\prime} \CH \ket{\Omega} = \cases{ N_c\  f(\Omega),&$\Omega
=\Omega^\prime$,\cr 0,&$\Omega \not= \Omega^\prime$,\cr}
}
where $f(\Omega)$ is a function of $\Omega$ which is a spin and flavor
singlet, \ie\ it is invariant under the transformation eq.~\VIiv.
An example of such a function is $f(\Omega) = f_0 + f_1 \epsilon^{\iota_1
\iota_2}\epsilon^{\alpha_1 \alpha_2}\Omega_{\iota_1 \alpha_1}
\Omega_{\iota_2 \alpha_2}$, where $f_0$ and $f_1$ are constants. Although the
above argument shows that matrix
elements of the Hamiltonian $\CH$ between states $\ket{\Omega}$ is of order
$N_c$, it does not imply that the states are degenerate since the order $N_c$
term
in the mass can be a function of $\Omega$, and there can also be terms of
order one in the mass.

A similar argument can be used to show that axial current
matrix elements between $\Omega$ states are of order $N_c$. The axial current
operator can be inserted on any one of the $N_c$ identical  quark lines. There
is no cancellation between the different possible insertions, since all the
quarks are identical. The quark axial current operator is a single quark
operator, which cannot affect the orthogonality relation for the
$\ket{\Omega}$'s, so
\eqn\VIvii{
\bra{\Omega^\prime} A^{ia} \ket{\Omega} = \cases{ N_c\
f^{ia}(\Omega),&$\Omega
=\Omega^\prime$,\cr 0,&$\Omega \not= \Omega^\prime$,\cr}
}
where $f^{ia}(\Omega)$ is a function of $\Omega$ which transforms like a
tensor with spin one and isospin one under the transformation eq.~\VIiv.
One such function is $f^{ia}(\Omega)= \Omega_{\iota_1 \alpha_1}\Omega_{\iota_2
\alpha_2} (\sigma^i)^{\iota_1 \iota_2} (\tau^a)^{\alpha_1 \alpha_2}$. Thus,
the matrix elements of the axial current between $\ket{\Omega}$
states is of order $N_c$.

Note that it is not possible to prove that the matrix elements of the spin
operator ${\bf
J}$ or isospin operator ${\bf I}$ are of order $N_c$ using this method, since
it is not possible to construct a function with spin one and
isospin zero (the quantum numbers of ${\bf J}$) or isospin one and spin zero
(the quantum numbers of ${\bf I}$) from $\Omega$, which is a $c$-number and
transforms as $(1/2,1/2)$ under spin $\otimes$ flavor. The only tensors that
can be constructed from the ${\rm m}^{\rm th}$ power of $\Omega$ are in the
totally symmetric tensor product $(1/2,1/2)^{\otimes m}$, which does not
contain $(1,0)$ or $(0,1)$.

\subsec{The Relation of $\ket{\Omega}$ States to Induced Representations}

Baryons annihilated by $B_\Omega$ are completely symmetric in spin
$\otimes$ flavor. The baryons have the quantum numbers of states in the
completely
symmetric tensor product  $(1/2,1/2)^{\otimes N_c}$, which are states with
$(J,I)$ equal to $(1/2,1/2)$, $(3/2,3/2)$, $\ldots$, $(N_c/2,N_c/2)$. In the
limit that \nclimit, it is easy to see that they correspond to the induced
representation discussed in the previous section with $K=0$, and $\pm$ =
$+$ if
the \nclimit\ limit is taken with $N_c$ even, and with $\pm=-$ if the limit is
taken with $N_c$ odd. The
connection between the states $\ket{\Omega}$ defined in this section and the
induced representation defined in the previous section is
\eqn\VIviii{
 X_0^{ia}\leftrightarrow \half \Tr \Omega^{-1}\tau^i \Omega \tau^a,
}
where $\tau$ are the Pauli matrices for $SU(2)$.\footnote{${}^\dagger$}{To be
precise, the $\Omega$ in eq.~\VIviii\ transforms as a ${\bf (\bar 2,2)}$
and is
obtained from the $\Omega$ in eq.~\VIiii\ which transforms as a ${\bf (2,2)}$
by raising one index using the $SU(2)$ epsilon symbol.}
 Note that the point $\Omega=1$
corresponds to the reference point $\bar X_0$ of the orbit. The quark picture
naturally gives orbits where $\bar X_0$ has all three eigenvalues equal since
the quark representation labels states by $\Omega$, which
transforms as $(1/2,1/2)$ under spin $\otimes$ isospin. The only object
which can be constructed out of $\Omega$ that transforms as $(1,1)$ like $X_0$
is
a bilinear in $\Omega$ of the form eq.~\VIviii\ which has all eigenvalues
equal.

A similar analysis can be done for baryons containing a finite number of
strange quarks as \nclimit. The baryon annihilation operator
\eqn\VIix{
q_{\iota_1 \alpha_1}^{(+)} q_{\iota_2 \alpha_2}^{(+)} \ldots
q_{\iota_{N_c-N_s} \alpha_{N_c-N_s} }^{(+)} \Omega_{\iota_1 \alpha_1}
\Omega_{\iota_2 \alpha_2}
\ldots
\Omega_{\iota_{N_c-N_s} \alpha_{N_c-N_s} } s^{(+)}_{k_1} s^{(+)}_{k_2}\ldots
s^{(+)}_{k_{N_s}} ,
}
annihilates baryons containing $N_s$ strange quarks. The
annihilation operator is completely symmetric under the exchange of the
strange
quark indices, so the spins of the $N_s$ strange quarks are combined to form a
completely symmetric state of spin $N_s/2$. These states correspond to the
induced representation discussed in the previous section with
$K=N_s/2$.

The results of this section strongly suggest that the lowest lying baryons
correspond to induced representations on orbits with $\bar X_0^{ia}$
proportional to $\delta^{ia}$, and with $2K$ equal to the number of strange
quarks in
the baryon. However, this identification is not rigorous because it has not
been
proved that the baryons annihilated by the operators eqs.~\VIiii\ and \VIix\
are the lowest lying
baryons. For example, the operator eq.~\VIiii\ with an additional
derivative on
one of the $q^{(+)}$'s, or with a $q^{(+)}$ replaced by a $q^{(-)}$ transforms
under a different representation of the spin-flavor algebra. One expects that
these states correspond to excited baryons, but this identification has not
been proved.

\newsec{Pion-Baryon Couplings}

The irreducible representations for the baryons constructed in Sect.~3
give the possible solutions to the pion-baryon consistency conditions.
The solutions can be classified by the quantum number $K$, where $2K$ is equal
to the number of strange quarks in the baryons.
Thus, the $K=0$ sector contains the strangeness zero baryons ($N$, $\Delta$);
the $K=1/2$ sector contains the strangeness $-1$ baryons ($\Lambda$,
$\Sigma$, $\Sigma^*$); the $K=1$ sector contains the $S=-2$ baryons
($\Xi$, $\Xi^*$); and the $K=3/2$ sector contains the $S=-3$ baryon $\Omega$.
Pion interactions only couple baryons within a given $K$ sector, since pions do
not carry strangeness.
The results of Sect.~3 allow us
to determine the pion couplings within each strangeness sector in terms of an
overall coupling constant $g(K)$
which can depend on $K$.  The pion couplings for each $K$ sector are first
derived at leading order in $1/N_c$.  The analysis is then extended to include
$1/N_c$ corrections.
It is important to remember that while the results in this section are for
baryons containing strange quarks, $SU(3)$ symmetry is not assumed and the
formul\ae\ are true irrespective of the size of the strange quark mass.  A
different approach which uses $SU(3)$ flavor symmetry is given in Sect.~9.

The pion-baryon couplings within a given $K$ tower can be computed using the
explicit formula eq.~\IVxviii\ for the baryon states $\ket{I\,I_3, J\,J_3
; K }$. The pion couplings are the matrix elements of $X_0^{ia}$, and
are given by
\eqn\VIIi{\eqalign{
&\bra{I^\prime\,I_3^\prime, J^\prime\,J_3^\prime; K}X_0^{ia}\ket{I\,I_3,
J\,J_3 ;K} =\cr
\noalign{\smallskip}
&\sqrt{\dim I \dim J \dim I^\prime \dim J^\prime\over  (\dim K)^2}
\clebsch{J^\prime}{J_3^\prime}{I^\prime}{m^\prime}{K}{k^\prime}
\clebsch{J}{J_3}{I}{m}{K}{k}\cr
\noalign{\smallskip}
&\int dg \,\int dh \ D_{I_3^\prime m^\prime}^{(I^\prime)}(g)\ D_{I_3
m}^{(I)}(h)^* \bra{\bar X_0,K,k^\prime} U_I(g)^\dagger X_0^{ia} U_I(h)
\ket{\bar X_0,K,k}.\cr
}}
The orthogonality of the states eq.~\IVxiv\
implies that the integrals over $g$ and $h$ in eq.~\VIIi\
collapse to a single integral over $g$. Using eq.~\IViii\ to evaluate
$U_I(g)^\dagger X_0^{ia} U_I(g)$ gives
\eqn\VIIii{\eqalign{
&\bra{I^\prime\,I_3^\prime, J^\prime\,J_3^\prime; K}X_0^{ia}\ket{I\,I_3,
J\,J_3 ; K} =\cr
\noalign{\smallskip}
&\sqrt{ \dim I \dim J \dim I^\prime \dim J^\prime\over  (\dim K)^2}
\clebsch{J^\prime}{J_3^\prime}{I^\prime}{m^\prime}{K}{k}
\clebsch{J}{J_3}{I}{m}{K}{k}\cr
\noalign{\smallskip}
&\int dg \ D_{I_3^\prime m^\prime}^{(I^\prime)}(g)\ D_{I_3 m}^{(I)}(g)^*\
D_{ab}^{(1)}(g)^* \ \bar X_0^{ib}, \cr
}}
where $D^{(1)*}=D^{(1)}$ since the representation is real, and  $\bar X_0^{ib}
= \delta^{ib}$ by the definition of the reference
point of the orbit eq.~\IVviii. The integral over $g$ can be done using
the identity on $D$ matrices,
\eqn\VIIiii{
\int dg \ D_{I_3^\prime m^\prime}^{(I^\prime)}(g)\ D_{I_3 m}^{(I)}(g)^*\
D_{ai}^{(1)}(g)^* ={1 \over {\dim {I^\prime}}}
\clebsch{I}{I_3}{1}{a}{I^\prime}{I_3^\prime}
\clebsch{I}{m}{1}{i}{I^\prime}{m^\prime}\ .
}
Substituting eq.~\VIIiii\ into eq.~\VIIii, and rewriting
three of the Clebsch-Gordan coefficients in terms of a
$6j$-symbol times a Clebsch-Gordan coefficient gives the result
\eqn\VIIiv{\eqalign{
&\bra{ I^\prime\, I_{3}^\prime, J^\prime\, J_{3}^\prime; K }
X_0^{ia}  \ket{ I \, I_{3}, J\, J_{3}; K }
= (-1)^{2J^\prime+J-I^\prime-K}\cr
&\qquad
\sqrt{ { \dim I } \, { \dim J } }
\sixj{1}{I}{I^\prime}{K}{J^\prime}{J}
\ \clebsch {I}{I_{3}}{1}{a}{I^\prime}{I^\prime_{3}}
\clebsch {J}{J_{3}}{1}{i}{J^\prime}{J^\prime_{3}}. \cr
}}
For the special case $K=0$, eq.~\VIIiv\ reduces to
\eqn\VIIv{
\bra{ I^\prime\, I_{3}^\prime, J^\prime\, J_{3}^\prime; 0 }
X_0^{ia}  \ket{ I\, I_{3}, J\, J_{3}; 0 }
= \sqrt{ { \dim J } \over { \dim J^\prime } }
\ \clebsch {I}{I_{3}}{1}{a}{I^\prime}{I^\prime_{3}}
\clebsch {J}{J_{3}}{1}{i}{J^\prime}{J^\prime_{3}},
}
which are the couplings of the $I=J$ tower containing the nucleon and
$\Delta$ baryons found in ref.~\dmi\ by explicit construction.
Eq.~\VIIiv\ determines the pion couplings in a given $K$ sector up to an
overall undetermined coupling constant $g(K)$. The coupling constants in
the different $K$ sectors are not related by pion-scattering, since
pions have zero strangeness and do not connect the different strangeness
sectors to each other.
Thus the final expression for the axial current matrix elements in the
large $N_c$ limit is
\eqn\VIIvi{\eqalign{
&\bra{ I^\prime\, I_{3}^\prime, J^\prime\, J_{3}^\prime; K }
A^{ia}  \ket{ I\, I_{3}, J\, J_{3} ; K }
= N_c \,g(K) (-1)^{2J^\prime+J-I^\prime-K}\cr
&\qquad
\sqrt{ { \dim I } \, { \dim J } }
\sixj{1}{I}{I^\prime}{K}{J^\prime}{J}
\ \clebsch {I}{I_{3}}{1}{a}{I^\prime}{I^\prime_{3}}
\clebsch {J}{J_{3}}{1}{i}{J^\prime}{J^\prime_{3}},
}}
where $g(K)$ is an unknown $K$ dependent normalization.

Eqs.~\VIIiv\ and \VIIvi\ are the $N_c \rightarrow \infty$ predictions for the
pion couplings of each $K$ tower.  The $1/N_c$ corrections to
the pion couplings in each $K$ tower can also be computed by
generalizing the method employed in ref.~\dmi\ to calculate $1/N_c$
corrections for the $K=0$ tower.  In order to discuss $1/N_c$
corrections to the pion couplings, it is useful to define an expansion
for the operator $X^{ia}$,
\eqn\VIIvii{
X^{ia}= X_0^{ia} + {1\over N_c} X_1^{ia} + {1\over N_c^2} X_2^{ia} + \ldots,
}
where $X_0^{ia}$ is the operator which describes the pion-baryon
couplings at leading order, and $X_n^{ia}$ are operators which arise as
$1/N_c^{n}$ corrections to the large $N_c$ limit. With this definition,
the consistency condition for pion-baryon scattering becomes
\eqn\VIIviii{
\left[ X^{jb}, X^{ia} \right] \ltap \CO\left( {1 \over N_c} \right),
}
since each pion-baryon vertex is order $\sqrt{N_c}$ and the scattering
amplitude is $\CO(1)$.  Eq.~\VIIviii\
implies the commutation relation $\left[X_0^{ia},
X_0^{jb} \right] =0$ which was used to obtain the leading order pion
couplings.

A consistency condition for the $1/N_c$ correction to the axial current
is obtained by considering the three-pion scattering process
$\pi^a(\omega_1+\omega_2) + B\rightarrow \pi^b(\omega_1)
+\pi^c(\omega_2)+B^\prime$ shown in \fig\fVi{The diagrams contributing to
$\pi+B\rightarrow \pi+\pi+B^\prime$ at leading order in $1/N_c$.}, where the
incident pion has
energy $\omega_1+\omega_2$, and the outgoing pions have energies
$\omega_1$ and $\omega_2$, respectively. Each pion-baryon vertex is of
order $\sqrt{N_c}$, so the scattering amplitude is order $N_c^{3/2}$.
The sum of the Feynman graphs in \fVi\ gives a total amplitude proportional to
\eqn\VIIix{
N_c^{3/2} {1\over \omega_1\omega_2(\omega_1+\omega_2)}
\left(\omega_1\left[X^{kc}, \left[X^{jb},X^{ia}\right]\right]+
\omega_2\left[X^{jb},\left[X^{kc}, X^{ia}\right]\right]\right).
}
The large $N_c$ power counting rules imply that the scattering amplitude
\VIIix\ is at most of order $1/\sqrt{N_c}$ in the large $N_c$ limit, which
means that the double commutator of three $X$'s must vanish at least
as fast as $1/N_c^2$,
\eqn\VIIx{
\left[X^{kc}, \left[X^{jb},X^{ia}\right]\right]\ltap \CO\left({1\over
N_c^2}\right).
}
At this order, the $1/N_c$ mass splittings of the baryons also
contribute to the scattering amplitude.  These terms have a different
energy dependence from eq.~\VIIix, however, and do not affect the double
commutator condition eq.~\VIIx.  They instead yield consistency conditions for
the
baryon mass splittings, which are presented in Sect.~10.
The constraint eq.~\VIIx\ restricts the form of the $1/N_c$ correction to the
pion couplings.  Eq.~\VIIx\ implies that $X_1^{ia}$ must satisfy
\eqn\VIIxi{
\left[X_0^{kc}, \left[X_0^{jb},X_1^{ia}\right]\right]
+\left[X_0^{kc}, \left[X_1^{jb},X_0^{ia}\right]\right]=0 \ .
}
Finding the complete set of solutions to eq.~\VIIxi\ is simplest in the
$\ket{X,K,k}$ basis rather than in a spin-isospin eigenstate basis of
baryon states.  In the $\ket{X,K,k}$ basis, the $X_1^{ia}$ are written as
functions of $X_0$, partial derivatives $\partial/\partial X_0$, and operators
$O_K$ acting on the $k$ indices. There are no operators $O_K$ for the
$K=0$ sector (strangeness zero baryons), so $X_1$ is only a function of
$X_0$ and $\partial/\partial X_0$, \ie\ a function of $X_0^{ia}$, $J^i$
and $I^a$. The commutator of $X_0$ with a polynomial in $X_0$, $J$ and
$I$ reduces the degree of the polynomial in $J$ and $I$ by one, since
$X_0$ satisfies the commutation relations \IIIiv. Thus, the
constraint eq.~\VIIxi\ implies that $X_1$ can be at most linear in $J$ or
$I$. The only solution for $X_1^{ia}$ in the $K=0$ tower is $X_1^{ia}$
proportional to $X_0^{ia}$ \dmi, since the
two possible operators which are linear in $J$ and $I$,
\eqn\VIIxii{
\epsilon_{ijk}\ X_0^{ja} J^k,\qquad \epsilon_{abc}\ X_0^{ib} I^c,
}
do not transform under time reversal in the same manner as an axial
current. Equivalently, the operators in eq.~\VIIxii\ are commutators of $J^2$
and $I^2$ with $X_0^{ia}$, and can be removed by phase redefinition of the
baryon states of order $1/N_c$. Because the $1/N_c$ correction to the axial
currents in the $K=0$
sector is proportional to $X_0^{ia}$, it can be reabsorbed into the overall
normalization factor $g(K)$. Thus there are no $1/N_c$ corrections to the
ratios of pion couplings in the $K=0$ sector \dmi. The Ademollo-Gatto theorem
implies that conserved charges are not renormalized to first order in symmetry
breaking. Large $N_c$ QCD has a contracted spin-flavor symmetry that is broken
by $1/N_c$ corrections. At first order in $1/N_c$, one can get a correction to
the axial currents proportional to the lowest order values because the axial
currents are not normalized by the commutation relations eq.~\IIIiii--\IIIv.
The ratios of axial couplings are determined by the contracted symmetry, and
are not renormalized at first order in symmetry breaking.

To solve eq.~\VIIxi\ for $K\not=0$ is more complicated.  Define
angular momenta of the light degrees of freedom and
the strange quarks of the baryon by $J_{ud}$ and $J_s$, respectively,
where $J = J_{ud} + J_s$. The angular
momentum $J_{ud}$ acts only on the $X_0$ variables in
$\ket{X_0,K,k}$,
\eqn\VIIxiii{
J_{ud}^i = -i\ \epsilon_{ij\ell}\ \delta_{ k^\prime k}\ X_0^{jc}{ \partial
\over \partial X_0^{\ell c}} ,
}
and the angular momentum of the strange quarks acts only on the $\ket{K,k}$
variables,
\eqn\VIIxiv{
J_s^i = T^{(K)i}_{k^\prime k}.
}
Any irreducible tensor operator $O_L$ with angular momentum $L$ acting on the
$\ket{K,k}$ variables is proportional to the totally symmetric and traceless
tensor product of $L$ $J_s^i$'s, by the Wigner-Eckart theorem. Thus any
operator
in a given $K$ sector can
be written as a product of $X_0^{ia}$, $I^a$, $J_{ud}^i$ and $J_s^i$,
with at most one factor of $J_{ud}$ or $I$ and at most $K$ factors of $J_s$.
The possible operators
can be simplified using the identities
\eqn\VIIxv{\eqalign{
X_0^{ia} X_0^{ib} &= \delta^{ab},\cr
X_0^{ia} X_0^{ja} &= \delta^{ij},\cr
\epsilon_{ijk}\ X_0^{ia} X_0^{jb} &= \epsilon_{abc} X_0^{kc},\cr
X_0^{ia} I^a &= J_{ud}^i,\cr
X_0^{ia} J_{ud}^i &= I^a.\cr
}}
In the $K=0$ sector, there were no possible operators with at most one
power of $J_{ud}$ or $I$ with the correct transformation properties to be an
axial current.
In the $K\not=0$ sector, there is an operator
\eqn\VIIxvi{
X_1^{ia}\propto J_s^i\ I^a
}
which satisfies eq.~\VIIxi\ since $X_0$ commutes with $J_s$, and has the
correct time-reversal properties. This term can be shown to be absent
by considering the process $\pi + B \rightarrow \pi + K + B^\prime$
involving two pions and a kaon.  This discussion is deferred to the next
section on kaon couplings.
Thus the only $1/N_c$ corrections to the axial currents are proportional
to $X_0$, and can be absorbed into a redefinition of $g(K)$ \dmi.  The
pion-baryon couplings including $1/N_c$ corrections
are still given by eq.~\VIIvi. The $1/N_c$ correction to $X^{ia}$ is
proportional to $X_0^{ia}$, so one finds that\footnote{${}^\dagger$}{This
commutator condition implies that the $\Delta$ contribution cancels the $g_A^2$
term in the Adler-Weisberger sum rule to two orders in $N_c$.}
\eqn\VIImmnni{
\left[X^{ia},X^{jb}\right] \ltap \CO\left({1\over N_c^2}\right),
}
which makes a contribution to the $\pi + B\rightarrow \pi + B^\prime$
scattering
amplitude
of order $1/N_c$. The order one contribution to the scattering amplitude
results from the $1/N_c$ correction to the intermediate baryon propagator due
to the baryon mass splittings, which are of order $1/N_c$.

The $1/N_c^2$ correction to the axial currents is constrained by the
four-pion scattering process $\pi + B \rightarrow \pi + \pi + \pi +
B^\prime$ which gives consistency conditions for the $1/N_c^2$
correction $X_2^{ia}$. The scattering amplitude for this process
contains terms of the form
$N_c^2\left[X,\left[X,\left[X,X\right]\right]\right]$ and
$N_c^2\left[\left[X,X\right],\left[X,X\right]\right]$, which must be of
order $1/N_c$, by the large $N_c$ power counting rules. Since
$\left[X,X\right]$ is of order
$1/N_c^2$, the second commutator condition is automatically
satisfied. The first condition  places a restriction on $X_2$,
\eqn\VIIxvii{
\left[ X_0^{\ell d}, \left[ X_0^{kc} , \left[ X_0^{jb} , X_2^{ia}
\right]\right] \right] + \left[ X_0^{\ell d}, \left[ X_0^{kc} , \left[
X_2^{jb} , X_0^{ia} \right]\right] \right] =0 \ .
}
The most general solution of eq.~\VIIxvii\ (and the analogous conditions
obtained by replacing pions by kaons) is that $X_2$ can have the form
$ J^i\, I^a $, $ J^i_s\, I^a $, $\{ I^2, X_0^{ia} \} $, and $\{ J^2,
X_0^{ia} \} $, as well as terms proportional to the lowest order
operator $X_0$ which can be reabsorbed into a redefinition of $g(K)$.

The coupling constant constant $g(K)$ depends on $K$ and $1/N_c$. The
coupling constant $g(K)$ for baryons in the $K$ sector has the form
\eqn\VIIxviii{
g(K) = c_0 + {1 \over N_c} c_1 K + {1\over N_c^2 } c_2 K^2
+ \CO\left({1\over N_c^3 }\right),
}
where the $K$-independent coefficients $c_i$ have expansions in powers of
$1/N_c$,
\ie\ the term of order one is $K$ independent, the term of order $1/N_c$
is at most linear in $K$, and the term of order $1/N_c^2$ is at most
quadratic in $K$. This form for $g(K)$ is derived in the next section. Thus,
$g(K)$ can be considered
to be a polynomial in $1/N_c$ and $K/N_c$. It is important to remember
that this
formula was obtained without using $SU(3)$ symmetry. Ratios of pion
couplings within a given $K$ tower are given by eq.~\VIIvi\ up to
corrections of order $1/N_c^2$, since $g(K)$ drops out in the ratio.
Ratios of pion-couplings between two different towers can have $1/N_c$
corrections due to the $K$ dependent term in $g(K)$ which is linear in
$K$ and of order $1/N_c$.

\newsec{Kaon Couplings}

The kaon couplings between baryons can be obtained by studying
kaon-baryon scattering in the large $N_c$ limit. Consistency
conditions for processes involving both pions and kaons also restrict
the form of the pion couplings discussed in the preceding
section. The results of this section are derived without assuming $SU(3)$
symmetry, and are valid irrespective of the size of the strange quark mass.
The analysis of kaon-baryon couplings is similar to the discussion of
heavy quark meson-baryon couplings in ref.~\eji.

The pion axial current matrix elements are of order $N_c$ in the large
$N_c$ limit, so that the pion-baryon vertex is of order $\sqrt{N_c}$.
The axial vector matrix elements of a strangeness changing current between
baryon states containing a finite number of strange quarks as $N_c \rightarrow
\infty$
are suppressed by a factor of $1/\sqrt{N_c}$ relative to the pion couplings.
Consider, for example, the $\Delta S=1$ transition matrix element
between a baryon containing a single strange quark, and a baryon
containing no strange quarks. The strangeness
changing axial current must be inserted on the strange quark line. The
baryon containing a strange quark is a linear superposition of states in
which the $n^{\rm th}$ quark is the $s$-quark ($1\le n\le N_c$), each with
amplitude $1/\sqrt{N_c}$. The net amplitude is of order $N_c$ (the
possible insertions on any of the quark lines) times $1/\sqrt{N_c}$
(the amplitude that the given quark line is an $s$-quark), \ie\ of order
$\sqrt{N_c}$. Thus the kaon-baryon vertex is of order one, since $f_K$
is of order $1/\sqrt{N_c}$. The constraints of the large $N_c$ limit of QCD on
the kaon
couplings are less powerful than for pions, because the kaon-baryon
vertex does not grow with $N_c$. Nevertheless, it is still possible to derive
the kaon couplings to leading order in the large $N_c$ limit.

A general kaon-baryon coupling is written as the baryon strangeness changing
axial current matrix element
\eqn\VIIIi{
\bra{B^\prime} \bar s \gamma^i\gamma_5 q^\alpha \ket{B} = \sqrt{N_c}\,
\left(Y^{i\alpha}\right)_{B^\prime B},
}
times the derivatively couplied kaon field $\partial^i \bar K^\alpha/f_K$,
where the index $\alpha$ represents the isospin flavor index of the kaon and
the index $i$ is a spin index.  This $p$-wave kaon carries spin one and
isospin
$1/2$.  The labels $B$ and $B^\prime$ denote baryons in the $K$ and $K+1/2$
sectors, respectively, since the absorption of a $\bar K$ adds a strange quark
to the baryon. A similar equation for the
matrix elements of $\bar q^\alpha \gamma^i\gamma_5 s$ defines the
hermitian conjugate matrix
$Y^{\dagger\, i \alpha}$ describing couplings of $K$'s to baryons. Since
$f_K \sim \sqrt{N_c}$ in the large $N_c$ limit, the baryon-kaon vertex
is $\CO(1)$.

Consider the scattering amplitude
for $\bar K^\alpha(\omega, {\bf k}) + B \rightarrow \pi^b(\omega, {\bf
k^\prime})+ B^\prime$, shown in \fig\VIi{A diagram
contributing to $\bar K + B \rightarrow \pi + B^\prime$. The thick line
is the $s$-quark.}, which turns a baryon $B$ into a baryon $B^\prime$ with
one additional strange quark.  The scattering amplitude for this process is
given by
\eqn\VIIIii{
\CA = -i \ { {{N_c}^{3/2}} \over {f_\pi f_K} } { {k^i k^{\prime j} } \over
\omega}
\left[g(K)\, X^{jb},  Y^{i\alpha}\right]_{B^\prime B},
}
where the coupling $g(K)$ must be retained in the commutator
because the amplitude depends on initial, final and intermediate baryons in
different $K$ sectors.  Since $f_{\pi,K} \sim \sqrt{N_c}$, the scattering
amplitude naively grows like $\sqrt{N_c}$.  Large $N_c$ power counting rules
imply that the amplitude should be order $1/\sqrt{N_c}$, so
the large $N_c$ consistency condition for $\bar K^\alpha +B\rightarrow
\pi^b +B^\prime$ scattering is
\eqn\VIIIiii{
\left[g(K)\, X^{jb},  Y^{i\alpha}\right]\ltap\CO\left({1 \over N_c
}\right).
}
As for the pion couplings $X$, define the $1/N_c$ expansion for the kaon
couplings $Y$ by
\eqn\VIiv{
Y^{ia}= Y_0^{i\alpha} + {1\over N_c} Y_1^{i\alpha} + \ldots.
}
Eq.~\VIIIiii\ implies that $Y_0^{i\alpha}$ must satisfy
\eqn\VIIIv{
\left[g(K)\, X_0^{jb}, Y_0^{i\alpha}\right]=0 \ .
}
It is important to remember that the kaon coupling changes the value of
$K$ by $1/2$.  Thus $g X_0$ in the two terms of the commutator eq.~\VIIIv\
refer
to the pion couplings in two different towers with $K$ values differing by
$1/2$.
The solution of eq.~\VIIIv\ is simple in the $\ket{X_0, K,k}$ basis.
Taking the matrix element of eq.~\VIIIv\ yields
\eqn\VIIIvi{
\bra{X_0^\prime,K+\frac 1 2,k^\prime}\left[g(K) X_0^{jb},
Y_0^{i\alpha}\right]\ket{X_0, K, k} =0 \ .
}
Inserting a complete set of states and using eq.~\IVi\ gives
\eqn\VIIIvii{
\left( g\left(K+\frac 1 2\right)\,X_0^{\prime \, jb} - g(K)\, X_0^{jb} \right)
\bra{X_0^\prime,K+ \frac 1 2,k^\prime}
Y_0^{i\alpha}\ket{X_0, K, k}=0 ,
}
which implies that $g(K)=g\left(K+ \frac 1 2\right)$ if $X_0=X_0^\prime$ and
that
\eqn\VIIIviii{
\bra{X_0^\prime,K+\frac 1 2,k^\prime}
Y_0^{i\alpha}\ket{X_0, K, k}=0 \qquad {\rm if}\ X_0\not= X_0^\prime.
}
The equality on $g(K)$ which is a consequence of the consistency
condition eq.~\VIIIv\ can be rewritten as
\eqn\VIIIix{
g(K) = g(0) + \CO\left({1\over N_c}\right),
}
which proves the assertion of the previous section that the order
one contribution to $g(K)$ is independent of $K$.

The constraint eq.~\VIIIviii\ on the kaon couplings implies that the operator
$Y_0$ does not change the value of the collective coordinate $X_0$.  Thus,
$Y_0$
can be written as a function of $X_0$ and operators acting on
$\ket{K,k}$, with no derivatives with respect to $X_0$. The matrix
element of $Y_0$ between general $X_0$ states is related to the matrix
element between states at the standard reference point
$\bar X_0^{ia} = \delta^{ia}$ by a group transformation,
\eqn\VIIIx{
\bra{X_g,K+\frac 1 2,k^\prime} Y_0^{i\alpha} \ket{X_g,K,k}
=D^{(1/2)}_{\alpha\beta}(g)\bra{\bar X_0,K+\frac 1 2,k^\prime} Y_0^{i\beta}
\ket{\bar X_0,K,k},
}
where $X_g$ is obtained from $\bar X_0$ by an isospin rotation $g$, and
$D^{(1/2)}_{\alpha\beta}$ is the rotation matrix in the spin-1/2
representation. The matrix element on the right-hand side of eq.~\VIIIx\
is determined by
considering the transformation properties of $Y_0$ under the little group
generated by ${\bf K}$. Since $Y_0$ has spin
one and isospin $1/2$, it transforms as a linear combination of irreducible
tensor operators with $\Delta K=1/2$ and $3/2$.  These operators must be
combined with
the states $\ket{K,k}$ and $\ket{K+\frac 1 2,k^\prime}$ in $K$-invariant
linear combinations. The state $\ket{K,k}$ can be considered to be the
completely symmetric tensor product of $2K$ strange quarks, each with
spin-1/2. The state $\ket{K+\frac 1 2,k^\prime}$ is then the completely
symmetric tensor product of $(2K+1)$ strange quarks. Any transition
operator between $\ket{K,k}$ and $\ket{K+\frac 1 2,k^\prime}$ can be written
in terms of products of creation and annihilation operators $a^\dagger_\alpha$
and
$a^\alpha$ which create and annihilate a strange quark with spin
$\alpha$. To make a transition from $K$ to $K+1/2$, the operator must
have one more $a^\dagger$ than $a$.
Any operator that transforms as $\Delta K=1/2$ or $\Delta K=3/2$ between $K$
and $K+1/2$ states is proportional to a
linear combination of $a^\dagger_\alpha$ and $a^\dagger_\alpha
a^\dagger_\beta a^\gamma$ by the Wigner-Eckart theorem. The first operator is
pure $\Delta K=1/2$, and the
second is a linear combination of $\Delta K=1/2$ and $\Delta K=3/2$.
Thus the most general form for
$Y_0^{i\alpha}$ at the standard point $\bar X_0$ of the orbit is
\eqn\VIIIxi{
Y_0^{i\alpha} = c(K)\,  a^\dagger_\lambda \left(
\sigma^i\right)^\lambda{}_\alpha
+ d(K)\, a^\dagger_\alpha \left(a^\dagger\sigma^i a \right),
}
where $c(K)$ and $d(K)$ are coefficients which depend on $K$. The
$\Delta K=1/2$ operator proportional to $c(K)$ preserves the $s$-quark spin
symmetry
of the baryons, whereas the operator proportional to $d(K)$ violates the
$s$-quark spin symmetry.

The $\Delta K=3/2$ operator is forbidden by a large $N_c$ consistency
condition
obtained from $\bar K^\alpha + B \rightarrow K^\beta + B^\prime$
scattering. Naively, the amplitude is of order one since each
kaon-baryon vertex is of order one. However, large $N_c$ quark counting
rules show that the amplitude is at most of order $1/N_c$, which leads
to the consistency condition
\eqn\VIIIxii{
\left[Y_0^{j\beta},Y_0^{i\alpha}\right]=0.
}
There is an important subtlety when one considers kaon-baryon
scattering. The quark counting rules show that  $\bar K^\alpha + B
\rightarrow K^\beta +B^\prime$ is of order $1/N_c$, but the process
$K^\alpha + B \rightarrow K^\beta + B^\prime$ is of order one. Thus one
obtains the consistency condition eq.~\VIIIxii\ but the condition
$\left[Y_0^{j\beta},Y_0^{\dagger i\alpha}\right]=0$ {\sl is not
satisfied}.
The consistency condition eq.~\VIIIxii\ requires that the coefficients
$d(K)$ in eq.~\VIIIxi\ vanish, so only the $\Delta K=1/2$ amplitude is allowed
at leading order in $1/N_c$. Thus, kaon-baryon couplings respect baryon
$s$-quark spin symmetry to leading order in $1/N_c$, even though we have
not assumed that the strange quark is heavy. Note that this result does not
imply that
strange quark spin symmetry is a good symmetry of the theory. For
example, there is no reason to believe that the couplings of the $K^*$
to baryons are related to the couplings of the kaon to baryons by $s$-quark
spin symmetry.

Eq.~\VIIIxi\ can be restricted further by considering
the scattering process $K^\alpha + B \rightarrow K^\beta + K^\gamma +
B^\prime$, which
yields the constraint
\eqn\VIIIxiii{
\left[Y^{k \gamma},\left[Y^{j\beta},Y^{\dagger
i\alpha}\right]\right]\ltap\CO\left({1 \over N_c} \right)\ .
}
Eq.~\VIIIxiii\ implies that $c(K)=c(0)$, a constant independent of $K$,
so that
\eqn\VIIIxiv{
Y_0^{i\alpha} = c \,  a^\dagger_\lambda \left(
\sigma^i\right)^\lambda{}_\alpha
}
is determined up to an overall normalization constant $c$.

Kaon couplings for baryon states of definite spin and isospin can now
be computed using eqs.~\IVxviii, \VIIIx\ and \VIIIxiv.  The matrix element
eq.~\VIIIx\ becomes
\eqn\VIIIxv{\eqalign{
\bra{X_g,K+\frac 1 2,k^\prime} Y_0^{i\alpha} \ket{X_g,K,k}
&=c \sqrt{\dim K}\ D^{(1/2)}_{\alpha\beta}(g) \cr
&\clebsch {\frac 1 2}{\beta} 1 i {\frac 1 2}\gamma
\clebsch {\frac 1 2} \gamma K k {K+\frac 1 2}{k^\prime},\cr
}}
where the equation
\eqn\VIIIxvi{
\bra{\bar X_0,K+\frac 1 2,k^\prime} a^\dagger_\beta  \ket{\bar X_0,K,k}
=\sqrt{2(K+\frac 1 2)}
\clebsch {\frac 1 2}{\beta} K k {K+\frac 1 2}{k^\prime}
}
has been used.
Using the definition of isospin and spin states eq.~\IVxviii\ yields after
considerable manipulation
\eqn\VIIIxvii{\eqalign{
&\bra{I^\prime\, I_3^\prime, J^\prime\, J_3^\prime; K^\prime} Y_0^{i\alpha}
\ket{I \,I_3, J\, J_3; K}
=c \sqrt{ \dim I \dim J \dim K \dim K^\prime }\cr
&\qquad
\ninej{\frac 1 2} 1 {\frac 1 2} I J K {I^\prime} {J^\prime} {K^\prime}
\clebsch I {I_3} {\frac 1 2} \alpha {I^\prime} {I_3^\prime}
\clebsch J {J_3} 1 i {J^\prime} {J_3^\prime},
\cr
}}
where the quantity in curly braces is the $9j$ symbol, and
$K^\prime=K+1/2$. This equation determines all the kaon coupling ratios to
leading order in $1/N_c$, without assuming $SU(3)$ symmetry.

One can also consider the scattering processes $\pi + B \rightarrow \pi+
K + B^\prime$, $\bar K + B \rightarrow K + \pi + B^\prime$, and
$\bar K + B \rightarrow K + K + B^\prime$ which give the
consistency conditions
\eqn\VIIIxviii{
\left[ Y^{k\gamma},\left[ X^{jb}, X^{ia}\right]\right]=\CO\left({1 \over
N_c^2}\right),
}
\eqn\VIIIxix{
\left[ X^{kc},\left[ Y^{j\beta}, Y^{i\alpha}\right]\right]=\CO\left({1 \over
N_c^2}\right),
}
and
\eqn\VIIIxx{
\left[ Y^{k\gamma},\left[ Y^{j\beta},
Y^{i\alpha}\right]\right]=\CO\left({1 \over
N_c^2}\right),
}
respectively.
These consistency conditions can be used to show that the term in $g(K)$
in eq.~\VIIxviii\ of order $1/N_c$ is at most linear in $K$ and that
$X_1^{ia}$ does not contain terms of the form $J_s I$, as was stated in the
previous section.

\newsec{$\eta$ Couplings}

The matrix elements of the $T^8$ axial current between baryon states
containing finitely many strange quarks as \nclimit\ is of order one, so
that the $\eta$ couplings are of order $1/\sqrt{N_c}$ with respect to
the kaon couplings, and order $1/N_c$ with respect to the pion
couplings. As for the pions and kaons, define the $\eta$ couplings by
the axial vector current
\eqn\IXi{
\bra{B^\prime} \bar q \gamma^i\gamma_5 T^8 q \ket{B} =
\left(Z^{i}\right)_{B^\prime B},
}
times the derivatively coupled $\eta$ field $\partial^i \eta / f_\eta$, where
$Z^i$ is a matrix with spin one and isospin zero. The $\eta$-baryon
vertex is order $1/\sqrt{N_c}$. The first non-trivial constraint on
$Z^i$ comes from the process $\eta+B\rightarrow \pi+\pi +B^\prime$, with
amplitude proportional to
\eqn\IXii{
N_c^{1/2} \left[X^{ia},\left[X^{jb},Z^k\right]\right] \ .
}
Large $N_c$ power counting rules require that the amplitude be order
$1/\sqrt{N_c}$, so that
\eqn\IXiii{
\left[X_0^{ia},\left[X_0^{jb},Z_0^k\right]\right]=0 \ .
}
A similar argument using the processes  $\eta+B\rightarrow \pi+K
+B^\prime$ and  $\eta+B\rightarrow K+K +B^\prime$ gives the constraints
\eqn\IXiv{
\left[X_0^{ia},\left[Y_0^{jb},Z_0^k\right]\right]=0 \ ,
}
and
\eqn\IXv{
\left[Y_0^{ia},\left[Y_0^{jb},Z_0^k\right]\right]=0 \ .
}
The solution to these equations is
\eqn\IXvi{
Z_0^i = a J^i + b J_s^i
}
where $a$ and $b$ are constants independent of $K$ to leading order in
$1/N_c$. The ratios of the different $\eta$ couplings are not completely
determined even at leading order, since they depend on the unknown ratio
$a/b$.

\newsec{Meson-Baryon Couplings in the $SU(3)$ Limit and $F/D$}

The previous three sections analyzed the pion, kaon and $\eta$ couplings
of the baryons without assuming $SU(3)$ symmetry. In this section, these
couplings are studied in the limit of $SU(3)$ symmetry using tensor
methods \ref\amunpub{A.V.~Manohar, unpublished}\ref\BSWcan{J.~Bijnens,
H.~Sonoda, and M.B.~Wise, Can. J. Phys 64 (1986) 1}\ref\kaplan{D.B.~Kaplan and
I.~Klebanov, \np{335}{1990}{45}}.
The large $N_c$
results derived earlier allow us to compute the $F/D$ ratio for the
baryon axial currents for $N_c=3$ to order $1/N_c$. They also
constrain the form of non-analytic chiral logarithmic corrections to the
axial currents.

The spin-1/2 baryons in the large $N_c$ limit transform according to the
$SU(3)$ tensor
\eqn\Xone{
\CB^i_{j_1 j_2 \ldots j_\nu}
}
with one upper and $\nu$ completely symmetric lower indices, where
$N_c=2\nu+1$. For
$N_c=3$, eq.~\Xone\ reduces to the usual baryon octet tensor with one
upper and
one lower index. The
spin-3/2 baryons transform according to the tensor
\eqn\Xii{
\CT^{i_1 i_2 i_3}_{j_1 j_2 \ldots j_{\nu-1}}
}
which is completely symmetric in its three upper and $(\nu-1)$ lower indices.
For $N_c=3$, eq.~\Xii\ reduces to the baryon decuplet with
three upper indices and no lower indices. Throughout this section, the
spin-1/2
baryons are referred to as octet baryons, and the spin-3/2 baryons as decuplet
baryons, even though these
are the dimensions of the representations only for $N_c=3$. What makes the
$SU(3)$ analysis subtle is that the form
of the $SU(3)$ tensor changes with $N_c$. In this section, these
``representation effects'' are eliminated in order to extrapolate the
large $N_c$ results consistently to $N_c=3$.

The most general meson couplings to the baryon octet in the
$SU(3)$ limit are given by the two possible invariants
\eqn\Xiii{
\CM\ \bar \CB_a^{b_1 b_2 \ldots b_\nu}\ \left(T^A\right)^a{}_c
\ \CB^c_{b_1 b_2 \ldots b_\nu}+
\CN\ \bar \CB_a^{c b_2 \ldots b_\nu}
\ \CB^a_{d b_2 \ldots b_\nu}
\ \left(T^A\right)^d{}_c,
}
where $T^A$ is the $SU(3)$ octet matrix corresponding to a meson of type A.
These invariants  reduce to the two invariants $\Tr \bar\CB
T^A \CB$ and $\Tr \bar \CB \CB T^A$ for $N_c=3$ with coefficients $(D+F)$
and $(D-F)$, respectively. Similarly, the meson-decuplet-octet
coupling is given in terms of a single $SU(3)$ invariant tensor
\eqn\Xiv{
\CL\ \bar \CT_{\alpha \mu \nu}^{b_1 b_2 \ldots b_{\nu-1}}
\ \left(T^A\right)^\mu{}_\beta
\ \CB^\nu_{\gamma b_1 b_2 \ldots b_{\nu-1}}\ \epsilon^{\alpha\beta\gamma}.
}
The components of the baryon tensors $\CB$ and $\CT$ are \amunpub\BSWcan\kaplan

\noindent for the proton:
\eqn\Xv{
\CB^1_{3 3 \ldots 3} = 1,
}
for the neutron:
\eqn\Xvi{
\CB^2_{3 3 \ldots 3} = 1,
}
for the $\Sigma^+$:
\eqn\Xvii{
\CB^1_{233\ldots 3} = {1\over
\sqrt \nu},
}
for the $\Sigma^-$:
\eqn\Xviii{
\CB^2_{133\ldots 3} = -{1\over
\sqrt \nu},
}
for the $\Sigma^0$:
\eqn\Xix{
\CB^1_{133\ldots 3} = -\CB^2_{233\ldots 3} = -{1\over\sqrt {2\nu}},
}
for the $\Lambda$:
\eqn\Xx{
\CB^1_{133\ldots 3} =
\CB^2_{233\ldots 3}= {1\over
\sqrt {4+2\nu}},\qquad
\CB^3_{3 3 \ldots 3}=-{2\over \sqrt{4+2\nu}},
}
for the $\Xi^0$:
\eqn\Xxi{
\CB^3_{2 3 3 \ldots 3}=
-\bfrac32\ \CB^2_{2 2 3 \ldots 3}  =
-3\ \CB^1_{123\ldots 3}=\sqrt{\bfrac{3}{{\nu(\nu+2)}}},
}
for the $\Delta^{++}$:
\eqn\Xxii{
\CT^{111}_{33\ldots 3} =1,
}
for the $\Sigma^{*+}$:
\eqn\Xxiii{
\bfrac43\ \CT^{111}_{133\ldots3}=
4\ \CT^{112}_{233\ldots3}=
-\CT^{113}_{333\ldots3}=
=\sqrt{\bfrac{4}{3( \nu + 3)}},
}
for the $\Xi^{*0}$:
\eqn\Xxiv{\eqalign{
&\bfrac12\ \CT^{133}_{333\ldots3}=
-\bfrac34\ \CT^{113}_{133\ldots3}=
-\bfrac32\ \CT^{123}_{233\ldots3}=\cr
\noalign{\smallskip}
&\CT^{111}_{113\ldots3}=
3 \ \CT^{112}_{123\ldots3}= 3\ \CT^{122}_{223\ldots3}=
\sqrt{{\bfrac{1}{(\nu+2)(\nu+3)}}},
}}
for the $\Omega^-$:
\eqn\Xxv{
\CT^{333}_{333\ldots3}=
-2\ \CT^{331}_{133\ldots3}=-2\ \CT^{332}_{233\ldots3}=\sqrt{{2\over3\nu-1}}.
}
The symmetry of the tensors $\CB$ and $\CT$ determines the other
non-zero components. For example, the $\Sigma^+$ tensor has
$\CB^1_{233\ldots3} = \CB^1_{323\ldots3}=
\CB^1_{332\ldots3}=\CB^1_{333\ldots2}= 1/\sqrt \nu$, \etc\
The coefficients of $\CM$, $\CN$, and $\CL$ obtained by using
eqs.~\Xv--\Xxv\ in eqs.~\Xiii\ and~\Xiv\ are
given in Tables~1 and~2, respectively.
\bigskip
\centerline{
\vbox{\tabskip=0pt\offinterlineskip
\def\space{height4pt&\omit&\omit&&\omit&&\omit&\cr}
\def\tablerule{\noalign{\hrule}}
\halign {\vrule# & \hfil $\quad #$ & $\rightarrow #\quad $ \hfil &
\vrule # & \hfil $\quad # \quad $ \hfil & \vrule # & \hfil $\quad # \quad $
\hfil & \vrule #\cr
\tablerule\space
& \multispan2 Amplitude && \CM && \CN &\cr\space\tablerule\space
& p & n \pi^+ && 1 && 0 &\cr\space\tablerule\space
& \Sigma^+ & \Sigma^0 \pi^+ && {1\over\sqrt{2}} && -{1\over \sqrt {2}
\,\nu} &\cr\space\tablerule\space
& \Sigma^+ &\Lambda \pi^+ && {\nu \over \sqrt{\nu(2\nu+4)}} &&{1 \over
\sqrt{\nu(2\nu+4)}} &\cr\space\tablerule\space
& p & \Lambda K^+ && -{2\over \sqrt{2\nu+4}} && {1\over\sqrt{2\nu+4}}
&\cr\space\tablerule\space
& p & \Sigma^0 K^+ && 0 &&- {1\over \sqrt{2\nu}} &\cr\space\tablerule\space
& p & p \eta && 1 && -2 &\cr\space\tablerule\space
& \Sigma^+ & \Sigma^+ \eta && 1 && -2 + {3\over\nu} &\cr\space\tablerule
\space
& \Lambda & \Lambda \eta && {\nu-4\over  \nu +2} && - {2\nu+1 \over
\nu +2} &\cr\space\tablerule
}}}
\smallskip
\centerline{Table 1}
\vskip1truein
\centerline{
\vbox{\tabskip=0pt\offinterlineskip
\def\space{height4pt&\omit&\omit&&\omit&\cr}
\def\tablerule{\noalign{\hrule}}
\halign {\vrule# & \hfil $\quad #$ & $\rightarrow # \quad$ \hfil &
\vrule # & \hfil $\quad # \quad$ \hfil & \vrule # \cr
\tablerule\space
& \multispan2 Amplitude && \CL &\cr\space\tablerule\space
& \Delta^{++} & p \pi^+ && 1 &\cr\space\tablerule\space
& \Sigma^{*+} & \Sigma^0\pi^+ && -(\nu+1) \sqrt{{1\over 6 \nu (\nu+3)}}
&\cr\space\tablerule\space
& \Sigma^{*+} & \Lambda \pi^+ && \sqrt{{2 (\nu+2)\over 3(\nu+3)}}
&\cr\space\tablerule
}}}
\smallskip
\centerline{Table 2}
\bigskip

The octet-octet and decuplet-octet meson couplings in the $SU(3)$ limit
are obtained using the Clebsch-Gordan coefficients computed using tensor
methods in Tables~1 and 2, and the unknown coefficients $\CM$, $\CN$ and
$\CL$. The large $N_c$ analysis of the preceding
sections determines the ratios of pion couplings in a given $K$ sector
up to corrections of order $1/N_c^2$, and can be used to determine the
ratio $\CN/\CM$ to order one and $\CL/\CM$ to order $1/N_c$.
What makes the
determination of $\CN/\CM$ possible is that the $K=1/2$ baryons states
contain two different isospin states in the octet, the
$\Lambda$ and $\Sigma$. The ratio of the
$\Sigma\rightarrow\Sigma\pi$ coupling to the
$\Sigma\rightarrow\Lambda\pi$ coupling is known to order $1/N_c$,
\eqn\Xxvi{
{\Sigma^+\rightarrow\Sigma^0\pi^+\over  \Sigma^+\rightarrow\Lambda\pi^+}
= {  \sqrt 6 \sixj 1 1 1 {\frac 1 2}{\frac 1 2}{\frac 1 2} \clebsch 1 0
1 1 1 1
\over
\sqrt 2 \sixj 1 0 1 {\frac 1 2}{\frac 1 2}{\frac 1 2} \clebsch 0 0 1 1 1 1}
= 1 + \CO\left(\bfrac 1 {N_c^2}\right),
}
where we have used eq.~\VIIvi\ for the pion couplings. The $SU(3)$ value for
this ratio is obtained from Table~1 to be
\eqn\Xxvii{
{\Sigma^+\rightarrow\Sigma^0\pi^+\over  \Sigma^+\rightarrow\Lambda\pi^+}
=\sqrt{1 + {2\over\nu}}\left({{\nu \CM - \CN}\over {\nu \CM + \CN} }\right)\ .
}
Expanding eq.~\Xxvii\ in a power series in $1/N_c$, and comparing with
eq.~\Xxvi\ gives the large $N_c$ prediction
\eqn\Xxviii{
{\CN\over \CM} =\bfrac12 + {\alpha\over N_c}+ \CO\left({1\over N_c^2}\right),
}
where we have denoted the
(unknown) coefficient of the $1/N_c$ term in the ratio by $\alpha$, because it
will
be needed later in this section.
An inspection of Table~1 shows that knowing $\CN/\CM$ to order one is
sufficient to determine all the octet-pion couplings to order $1/N_c$,
since the coefficient of $\CN$ is suppressed by $1/N_c$ relative to that
of $\CM$. The ratio of pion couplings for $\Delta^{++} \rightarrow p
\pi^+$ to $p \rightarrow n \pi^+$ or the ratio of
$\Sigma^{*+}\rightarrow\Sigma^0\pi^+$ to
$\Sigma^+\rightarrow\Lambda\pi^+$
can be used to determine the ratio $\CL/\CM$ to order $1/N_c$,
\eqn\Xc{
{\CL\over \CM} = {\sqrt{3}\over 2}+ \CO\left({1\over N_c^2}\right).
}
(The normalization of $\CL$ relative to $\CM$ depends on how one
normalizes the
spin invariants for $\CB$ and $\CT$. Eq.~\Xc\ is derived assuming that
the spin invariants for $\CT \rightarrow \CB \pi$ and $\CB\rightarrow
\CB\pi$ are normalized to be equal to their respective spin
Clebsch-Gordan coefficients.)

The large $N_c$ results obtained in the previous sections are consistent
with $SU(3)$ symmetry. For example, the ratio
\eqn\Xxix{
{\Sigma^{*+}\rightarrow\Sigma^0\pi^+\over  \Sigma^{*+}\rightarrow\Lambda\pi^+}
= {\sqrt 6 \sixj 1 1 1 {\frac 1 2}{\frac 3 2}{\frac 1 2} \clebsch 1 0 1 1 1 1
\over
\sqrt 2 \sixj 1 0  1 {\frac 1 2}{\frac 3 2}{\frac 1 2} \clebsch 0 0 1 1 1
1}=-\bfrac12+\CO\left(\bfrac 1 {N_c^2}\right),
}
is obtained using the large $N_c$ result eq.~\VIIvi, and is valid for any
value of the $s$-quark mass. The same ratio is obtained using
$SU(3)$ tensors and the Clebsch-Gordan coefficients in
Table~2. There is only a single $SU(3)$ invariant amplitude for
decuplet-octet pion couplings, so the coupling $\CL$ drops out of the
ratio to give
\eqn\Xxx{
{\Sigma^{*+}\rightarrow\Sigma^0\pi^+\over
\Sigma^{*+}\rightarrow\Lambda\pi^+}=
-(\nu+1) \sqrt{ {3\over 4 \nu(\nu+2)}} ,
}
which is valid in the $SU(3)$ limit, for any $N_c$. Expanding eq.~\Xxx\ gives
\eqn\Xmmnni{
{\Sigma^{*+}\rightarrow\Sigma^0\pi^+\over
\Sigma^{*+}\rightarrow\Lambda\pi^+}=
-\bfrac12+\CO\left(\bfrac 1 {N_c^2}\right) ,
}
since there is no $1/N_c$ term in the expansion of the Clebsch-Gordan
coefficient ratio in eq.~\Xxx.

In Sect.~6, the pion couplings of the different $K$ sectors were
computed up to an overall coupling constant $g(K)$ which had an expansion
of the form eq.~\VIIxviii. In the $SU(3)$ limit, the coefficient of the
term linear
in $K$ can be determined. Consider the ratio
\eqn\Xxxi{
{\Delta^{++}\rightarrow p\pi^+\over  \Sigma^{*+}\rightarrow\Sigma^0\pi^+}
={g(0)\over g(\frac 1 2)} {\sqrt 4 \bigsixj {\frac 3 2}{\frac 1 2} 1
{\frac 1 2}{\frac 3 2} 0 \bigclebsch {\frac 1 2}
{\frac 1 2}  1 1 {\frac 3 2} {\frac 3 2} \over
\sqrt 6 \sixj 1 1 1 {\frac 1 2}{\frac 3 2}{\frac 1 2} \clebsch 1 0 1 1 1 1}=
-\sqrt{6}\,{g(0)\over g(\frac 1 2)}
}
which follows from eq.~\VIIvi. In the $SU(3)$ limit, the same ratio can
be evaluated using
the Clebsch-Gordan coefficients in Table~2 in terms of no unknowns since
the coupling $\CL$ cancels in the ratio,
\eqn\Xxxii{
{\Delta^{++}\rightarrow p\pi^+\over  \Sigma^{*+}\rightarrow\Sigma^0\pi^+}
=-\sqrt 6\left(1+{1\over N_c}\right) + \CO\left(\bfrac 1{N_c^2}\right).
}
Comparing eq.~\Xxxi\ with eq.~\Xxxii\ gives
\eqn\Xxxiii{
{g(0)\over g(\frac 1 2)} = 1 + {1\over N_c}+\CO\left(\bfrac 1{N_c^2}\right),
}
so that
\eqn\Xxxiv{
{g(K)\over g(0)} = 1- {2K\over N_c}+\CO\left(\bfrac 1{N_c^2}\right).
}
in the $SU(3)$ limit.

The same result eq.~\Xxxiv\ also can be obtained using only the octet
couplings. The ratio
\eqn\Xxxv{
{p\rightarrow n\pi^+\over  \Sigma^{+}\rightarrow\Lambda\pi^+}
=- {g(0)\over g(\frac 1 2)}
{ \sqrt 4 \bigsixj {\frac 1 2} {\frac 1 2} 1 {\frac 1 2}{\frac 1 2} 0
\bigclebsch {\phantom{-}\frac 1 2} {-\frac 1 2} 1 1 {\frac 1 2}
{\frac 1 2} \over
\sqrt 2 \sixj 1 0 1 {\frac 1 2}{\frac 1 2}{\frac 1 2} \clebsch 0 0 1 1 1
1}= \sqrt 2 \,{g(0)\over g(\frac 1 2)}
}
using eq.~\VIIvi.  Evaluation of this ratio using
the Clebsch-Gordan
coefficients in Table~1 and eq.~\Xxviii\ for $\CN/\CM$ gives
eq.~\Xxxiii. Thus, all the pion-couplings
can be determined consistently to order $1/N_c$ in the $SU(3)$ limit.

The $F/D$ ratio for the pion couplings can now be determined to order
$1/N_c$, by extrapolating the large $N_c$ results to $N_c=3$.
Different extrapolations lead to different values for $F/D$, but
the differences are of order $1/N_c^2$. The extraction of $F/D$ for
$N_c=3$ from the large $N_c$ results is tricky because the baryon $SU(3)$
representation is a function of $N_c$, a subtlety that was not present
for $SU(2)$ representations. There are $1/N_c$ effects in the pion couplings
because the
baryon representations change with $N_c$, which leads to $1/N_c$
corrections in the Clebsch-Gordan coefficients of Tables~1 and 2. These
``group-theoretic'' $1/N_c$ representation corrections are completely
calculable, and must be eliminated before one can extrapolate $F/D$ to
$N_c=3$ with an accuracy of $1/N_c^2$.
One extrapolation method is to
equate the ratio
\eqn\Xxxvi{
{\Sigma^+\rightarrow\Sigma^0\pi^+\over
\Sigma^+\rightarrow\Lambda\pi^+}=1 + \CO\left({1\over N_c^2}\right),
}
computed in large $N_c$ with its value in terms of $D$ and $F$ at
$N_c=3$,
\eqn\Xxxvii{
{\Sigma^+\rightarrow\Sigma^0\pi^+\over  \Sigma^+\rightarrow\Lambda\pi^+}
= {\sqrt{3} F\over D},
}
which yields
\eqn\Xxxviii{
{F\over D} = {1\over \sqrt 3 }+ \CO\left({1\over N_c^2}\right) = 0.58,
}
for $N_c=3$.
Another method is to recall that the non-relativistic quark model
values for the ratios of pion couplings differ from large $N_c$ QCD at
order $1/N_c^2$. Thus the quark model prediction for $F/D$ at $N_c=3$ is
accurate to $1/N_c^2$, which implies
\eqn\Xxxix{
{F\over D} = {2\over 3 }+ \CO\left({1\over N_c^2}\right) = 0.67 \ .
}
A third method is to use the Skyrme model predictions \bsw\ for the ratio of
pion couplings at $N_c=3$, which also differs from large $N_c$ QCD at
order $1/N_c^2$,
\eqn\Xxxx{
{F\over D} = {5\over 9 }+ \CO\left({1\over N_c^2}\right) = 0.56 \ .
}
The difference of these three numbers is a consequence of $1/N_c^2$ effects;
each determination yields an acceptable value for $F/D$ at $N_c=3$.  The range
of the three values implies an $\CO(1/N_c^2)$ correction of about $0.1$.  In
the following,
we use
the quark model value $F/D=2/3$. Any of the three values agrees well with the
experimental value for the $F/D$
ratio, $0.58\pm0.04$ \ref\jm{R.L.~Jaffe and A.V.~Manohar, \np{337}
{1990} {509}}.
The ratio of the decuplet-decuplet and
decuplet-octet pion couplings to $F$ and $D$ are also determined to order
$1/N_c$, and may be taken to be the non-relativistic quark
model values. In the notation of ref.~\ref\ejamaxial{E.~Jenkins and
A.V.~Manohar, \pl{255} {1991} {558}, \pl{259} {1991} {353}}, the decuplet-octet
pion coupling $\CC=2D$, and the decuplet-decuplet pion coupling
$\CH=3D$. These results are known to agree with the experimental data
\ejamaxial\ref\bssaxial{M.N.~Butler, M.J.~Savage and R.P.~Springer, \pl{304}
{1993}{353}}.

One can also use ratios such as
\eqn\Xxxxi{
{p\rightarrow n\pi^+\over  \Sigma^{+}\rightarrow\Lambda\pi^+}
}
involving states in different $K$ towers to determine the $F/D$ ratio. In
this case, it is important to remember that there are $1/N_c$ terms in the
ratios of $g(K)$. These terms are purely group-theoretical in nature,
and arise
because the size of the weight diagram changes as a function of $N_c$. Since
they are only group-theoretic, they are calculable (as was done in
eq.~\Xxxiv), and can be eliminated so
that the extrapolation to $N_c=3$ is valid to order $1/N_c$. This procedure
yields
a value for $F/D$ at $N_c=3$ which is consistent with
eqs.~\Xxxviii--\Xxxx\ up to order $1/N_c^2$.

The ratio $\CN/\CM$ is known to leading order in $1/N_c$, but it determines
the ratios of the pion couplings up to order $1/N_c^2$, since the coefficients
of
$\CN$ in Table~1 for the pion couplings are suppressed by $1/N_c$ relative to
those of $\CM$. The leading order value for $\CN/\CM$ also determines the
leading order values for the kaon couplings. It is a straightforward exercise
to verify that the kaon couplings are of order $1/\sqrt{N_c}$ relative to the
pions, and are given by eq.~\VIIIxvii. In the $SU(3)$ limit, one finds
$c=-\sqrt 6 \ g(0)$ by comparing the ratio
\eqn\Xmmi{
{p\rightarrow \Sigma^0 K^+\over p\rightarrow n\pi^+}=-{1\over
2\sqrt{N_c}}={c\over \sqrt{24 N_c}\ g(0)},
}
obtained using eq.~\Xxviii\ and the $SU(3)$ Clebsch-Gordan coefficients,
and by using eqs.~\VIIvi\ and \VIIIxvii. The
$\eta$-baryon couplings are of order $1/N_c$ relative to the pion-baryon
couplings, and depend on the unknown $1/N_c$ term $\alpha$ in $\CN/\CM$ in
eq.~\Xxviii. This lack of a prediction for the $\eta$ couplings is related to
the fact that the $\eta$ couplings in eq.~\IXvi\ depend on two operators
at leading order in $1/N_c$. In the $SU(3)$ limit, one finds that
\eqn\Xmmii{
a = -{4\sqrt 2 \over 3}\alpha\ g(0), \qquad
b = -{18\sqrt 2 \over 3} g(0).
}

The above analysis considered baryons with finite strangeness in the large
$N_c$ limit, for which the matrix elements of pion, kaon and $\eta$
axial currents are of order $N_c$, $\sqrt N_c$, and $1$, respectively. This
analysis
corresponds to working near the top of the $SU(3)$ weight diagram of
\fig\fVIIIi{The $SU(3)$ weight diagram for the spin-1/2 baryons for $N_c$
colors. The long edge of the weight diagram has $(N_c-1)$ states.}. For
$N_c=3$, however, the weight diagram contains only eight
states,
and the nucleon states are not far away from the the other two corners of the
weight diagram. One can therefore imagine other ways of extrapolating the
large $N_c$ results to $N_c=3$. For example, one can instead consider
states of definite $U$ spin or definite $V$ spin, and use these states to
extrapolate from $N_c \rightarrow \infty$ to
$N_c=3$. For $U$ spin, the $K^0$, $\bar K^0$ and $-\pi^0/2+\sqrt 3 \eta/2$
couplings are of order $N_c$, the $K^+$, $K^-$, $\pi^+$, $\pi^-$ couplings are
of order $\sqrt{N_c}$, and the $\sqrt 3 \pi^0/2+\eta/2$ coupling is of order
one.
Nevertheless, the $U$ spin or $V$ spin extrapolations give the same
$F/D$ ratio as the above procedure, up to errors of order $1/N_c^2$.

\subsec{Equal Spacing Rule for the Axial Couplings}

The pion-baryon coupling $g(K)$ is linear in $K$ at order $1/N_c$, as given in
eq.~\VIIxviii. This result was derived with no assumption of $SU(3)$ symmetry,
so it implies that $SU(3)$ breaking in the pion couplings is linear in $K$ to
order $1/N_c$. The $SU(3)$ breaking in the pion couplings can be extracted from
the pionic decays of the decuplet baryons to octet baryons.
The values of $\CC$ (defined in ref.~\ejamaxial) for $\Delta\rightarrow N\pi$,
$\Sigma^*\rightarrow \Lambda \pi$, $\Sigma^*\rightarrow\Sigma\pi$ and
$\Xi^*\rightarrow\Xi\pi$ are 1.8, 1.5, 1.5 and 1.3,
respectively~\ref\hungary{E.~Jenkins and A.V.~Manohar, {\it Baryon Chiral
Perturbation Theory}, in Proceedings of the Workshop on Effective Field
Theories of the Standard Model, ed. by U.~Meissner, (World Scientific, 1992)},
and should all be equal in the $SU(3)$ limit. The couplings clearly satisfy the
linearity constraint on $SU(3)$ breaking. In particular, the $SU(3)$ breaking
in the ratio  $\Sigma^*\rightarrow \Lambda \pi/\Sigma^*\rightarrow\Sigma\pi$ is
very small because both decays involve states in the $K=1/2$ sector. Since pion
couplings within a given $K$ tower are determined to order $1/N_c$, the $SU(3)$
breaking in $\CC$ between $\Delta\rightarrow N\pi$ and $\Sigma^*\rightarrow
\Lambda\pi$ must be the same as the $SU(3)$ breaking in the axial couplings for
the beta decays $n\rightarrow p e^-\bar\nu$ and $\Sigma \rightarrow \Lambda
e\nu$. This observation will allow a better determination of the $F/D$ ratio
from hyperon semileptonic decays, since $SU(3)$ breaking extracted from
decuplet-octet baryon pion couplings can be subtracted from the beta decay
couplings before performing the $SU(3)$ fit.

\subsec{Magnetic Moments}

The $SU(3)$ analysis of the baryon magnetic moments is similar to that of the
axial currents. The stability of the baryon magnetic moments under
renormalization, or equivalently, the large $N_c$ behavior of
pion-photoproduction, implies that at leading order, the baryon magnetic
moments must be proportional to the axial currents \dmi. This implies that
the $F/D$ ratio
for the baryon magnetic moments is also 2/3, in good agreement with the
experimental value of 0.72. The difference between the experimental values for
the $F/D$ ratios of the axial currents and magnetic moments is a $1/N_c^2$
correction. Thus the experimental data indicate that the $1/N_c^2$ correction
is
about 15--20\%, which indicates that the $1/N_c$ expansion (at least for these
quantities) is a reasonable expansion even for $N_c=3$. Model independent
relations for the baryon magnetic moments in the $1/N_c$ expansion are derived
in ref.~\ref\adkinsnappi{G.S.~Adkins and C.R.~Nappi, \np{249} {1985} {207}}\ in
the context of the Skyrme model.

\newsec{Baryon Masses}

The $1/N_c$ expansion restricts the form of the baryon mass spectrum. In this
section, we derive mass relations which are valid up to order $1/N_c$ without
imposing $SU(3)$ symmetry.  These mass relations constrain the form
of $SU(3)$ breaking in the baryon mass spectrum.

The mass of a baryon in
large $N_c$ has an expansion of the form
\eqn\XIIi{
M = N_c M_{0} + M_1 + {1 \over N_c} M_2 + \ldots \ ,
}
where the leading contribution to the mass is order $N_c$, so that the baryon
is infinitely heavy in the large $N_c$ limit.  Consistency conditions for the
baryon masses can be derived from the scattering amplitudes
considered in Sects.~6 and~7. Since baryon mass splittings are suppressed by
powers of $1/N_c$
relative to the leading mass $N_c M_{0}$, it is possible to expand the baryon
propagator about the static limit.  The static baryon propagator
is $i/k\cdot v$, where $v$ is the baryon velocity and $k$ is the residual
momentum of the baryon. Baryon mass
splittings change the baryon propagator to
\eqn\XIIii{
{i \over {k\cdot v - \Delta M}} \ ,
}
where $\Delta M$ is the mass difference of the intermediate and initial
baryons.
In the rest frame of the baryon, the propagator reduces to $i/(\omega -
\Delta M)$.  Consistency conditions on the pion-baryon scattering amplitudes
are
valid for arbitrary pion energy, provided the energy is held fixed
as $N_c\rightarrow\infty$.  When the pion energy is greater than $\Delta M$,
it is possible to expand the propagator eq.~\XIIii\ in a power series in
$\Delta M/(k \cdot v)$.  Since the terms in the expansion of the baryon
propagator have different energy dependences, each term in the
expansion of the propagator must separately satisfy the consistency
conditions. The leading term in the expansion of the propagator is the
lowest order propagator $i/k\cdot v$, which gives the consistency
conditions for the pion couplings. The next term in the expansion of the
propagator is $i\Delta M/(k\cdot v)^2$, which gives consistency
conditions for baryon masses.

The pion-baryon scattering amplitude for $\pi+B\rightarrow \pi+B^\prime$ is
naively of order $N_c$, but the quark counting rules imply that it is
at most of order one. This constraint was used to show that
$\left[X,X\right]$
is at most of order $1/N_c$, using the lowest order propagator. The
condition obtained by keeping the term proportional to $\Delta M$ in the
expansion of the propagator is
\eqn\XIIiii{
\left[ X^{jb}, \left[ X^{ia}, M \right] \right]\ltap\CO\left({1\over
N_c}\right),
}
which implies that
\eqn\XIIiv{\eqalign{
\left[ X_0^{jb}, \left[ X_0^{ia}, M_0 \right] \right] &= 0,\cr
\left[ X_0^{jb}, \left[ X_0^{ia}, M_1 \right] \right] &= 0.
}}
Similarly the constraint that the amplitude $\pi+B\rightarrow
\pi+\pi+B^\prime$ be at most of order $1/\sqrt{N_c}$ gives the constraint
\eqn\XIIv{
\left[ X^{kc}, \left[ X^{jb}, \left[ X^{ia}, M \right] \right] \right] \ltap
\CO\left({1\over N_c^2}\right),
}
which implies that $M_2$ satisfies
\eqn\XIIvi{
\left[ X_0^{kc}, \left[ X_0^{jb}, \left[ X_0^{ia}, M_2 \right] \right]
\right]=0.
}
A simpler (but equivalent) form of the above condition was derived in
ref.~\eji\ using chiral perturbation theory,
\eqn\XIIvii{
\left[ X_0^{ia}, \left[ X_0^{ia}, M_2 \right] \right] = {\rm constant} .
}
The solutions to eqs.~\XIIiv\ and~\XIIvi\ are that $M_0$ and $M_1$ are
independent of $I$ and $J_{ud}$, and $M_2$ is at most quadratic in
$I$ and $J_{ud}$.  Arbitrary dependence on $K$
and on $J_s$ is allowed.

Further restrictions on the form of $M$ are obtained by studying
scattering amplitudes involving kaons, since these processes constrain the
$K$ and $J_s$ dependence of $M$. Expanding out the intermediate propagator to
first order in $\Delta M$ in $\pi + B \rightarrow K + B^\prime$ and $\bar K + B
\rightarrow K + B^\prime$ scattering, and in $\bar K +B \rightarrow
\pi+\pi+B^\prime$,
$\bar K + B \rightarrow K + \pi+B^\prime$, and $\bar K + B \rightarrow K + K +
B^\prime$ scattering results in the
constraints
\eqn\XIIc{
\left[Y^{j\beta}, \left[ X^{ia}, M \right] \right]
 \ltap \CO\left({1\over N_c}\right),
}
\eqn\XIIviii{
\left[Y^{j\beta}, \left[ Y^{i\alpha}, M \right] \right]
 \ltap \CO\left({1\over N_c}\right),
}
and
\eqn\XIIxi{
\left[ X^{kc}, \left[ X^{jb}, \left[ Y^{i\alpha}, M \right] \right]
\right] \ltap
\CO\left({1\over N_c^2}\right),
}
\eqn\XIIx{
\left[ X^{kc}, \left[ Y^{j\beta}, \left[ Y^{i\alpha}, M \right] \right]
\right] \ltap
\CO\left({1\over N_c^2}\right),
}
\eqn\XIIix{
\left[ Y^{k\gamma}, \left[ Y^{j\beta}, \left[ Y^{i\alpha}, M \right] \right]
\right] \ltap
\CO\left({1\over N_c^2}\right).
}
Eq.~\XIIviii\ restricts $M_1$ to be at most linear in $K$, while
eq.~\XIIix\ restricts $M_2$ to be at most quadratic in $K$.  $M_0$ is
independent of $K$.

The most general solution of eqs.~\XIIiv--\XIIix\ is that $M$ has the form
\eqn\XIIxii{
M = N_c\ m_0 + m_1\ K + {1\over N_c} \left(m_2 \ I^2 + m_3\ J^2 + m_4 \
K^2\right),
}
where $m_i$ are constants independent of $K$ which have an expansion in
powers of $1/N_c$. (Note that the operator $J_s^2$ is equal to $K(K+1)$ and
that $J\cdot J_s$ can be written as a linear combination of $I^2$, $J^2$ and
$K(K+1)$.)
The baryon octet and decuplet each consist of four isospin multiplets.  These
eight baryon masses are parametrized by five mass parameters $m_i$ in
eq.~\XIIxii, so there are three mass relations amongst the octet and decuplet
masses which are valid up to corrections of order $1/N_c^2$.
These relations are any three of the following four mass relations,
\eqn\XIIxiii{
\frac 1 3 \left( \Sigma + 2 \Sigma^* \right) - \Lambda
= \frac 2 3 \left( \Delta - N \right),
}
\eqn\XIIxiv{
\Sigma^* - \Sigma = \Xi^* - \Xi,
}
\eqn\XIIxv{
\frac 3 4 \Lambda + \frac 1 4 \Sigma - \frac 1 2 \left( N + \Xi
\right) = - \frac 1 4 \left( \Omega - \Xi^* - \Sigma^* + \Delta
\right),
}
\eqn\XIIxvi{
\frac 1 2 \left( \Sigma^* - \Delta \right) - \left( \Xi^* -
\Sigma^* \right) + \frac 1 2 \left( \Omega - \Xi^* \right)=0 \ ,
}
each of which is valid including all terms of order $1/N_c$
in the baryon masses. {\sl These relations are true
irrespective of the mass of the strange quark, since they were derived
without assuming $SU(3)$ symmetry.} We will refer to the linear
combination of decuplet masses in eq.~\XIIxv\ as the equal spacing rule I,
and the combination in eq.~\XIIxvi\ as equal spacing rule II.
Eq.~\XIIxv\ relates the violation of the
Gell-Mann--Okubo formula to the violation of the equal spacing rule I for
the decuplet. The Gell-Mann--Okubo formula and the equal spacing rule I
are each violated at order $1/N_c$, but the difference of the two in
eq.~\XIIxv\ is only violated at order $1/N_c^2$. The equal spacing rule II
is only violated at order $1/N_c^2$.

One can characterize the deviations of the three $1/N_c$ mass relations by
$1/N_c^2$ operators.  Consistency conditions for $M_3$ are obtained from four
meson-baryon scattering.  There are three new operators which are first allowed
at order $1/N_c^2$,
\eqn\XIIwwi{
{1 \over N_c^2} M_3 = {1 \over N_c^2}\left( m_5 \, I^2 K + m_6 \, J^2 K + m_7
\, K^3 \right) \ ,
}
leaving no non-trivial relation amongst the eight octet and decuplet masses at
order $1/N_c^2$.  The eight masses can be used to determine the eight mass
parameters $m_0, \ldots, m_7$.  The corrections to eqs.~\XIIxiii--\XIIxvi\ are
$- (m_5 + m_6)/N_c^2$, $3 m_6/ 2 N_c^2$, $3(3 m_5 - m_7)/ 16 N_c^2$, and
$-3(m_5 + m_7)/8 N_c^2$, respectively.

One can also study the octet and decuplet relations at $N_c=3$ starting
from a $SU(3)$ symmetric Hamiltonian, and including a symmetry breaking
term proportional to $T^8$. In the symmetry limit, all the octet states
are degenerate, and all the decuplet states are degenerate, so that
eq.~\XIIxiii--\XIIxvi\ are satisfied.
The mass relations including terms of first
order in symmetry breaking are the Gell-Mann--Okubo formula for the
octet\footnote{${}^\dagger$}{This formula is true for arbitrary $N_c$.}
\eqn\XIIxvii{
\frac 3 4 \Lambda + \frac 1 4 \Sigma - \frac 1 2 \left( N + \Xi
\right) =0\ ,
}
and the equal spacing rule for the decuplet
\eqn\XIIxviii{
\Omega-\Xi^*=\Xi^*-\Sigma^*=\Sigma^*-\Delta \ .
}
These relations are violated by non-analytic terms of the form
$m_s^{3/2}$ and $m_s^2 \ln m_s$.  The equal
spacing rule II, however, has no $m_s^{3/2}$ and $m_s^2 \ln
m_s$ corrections~\ref\massej{E.~Jenkins, \np{368}{1992}{190}}.
Deviations from the mass relations can be characterized by combining the
expansion in the $SU(3)$-breaking parameter $m_s$ with the $1/N_c$ expansion,
\eqn\XIIxx{\eqalign{
\frac 1 3 \left( \Sigma + 2 \Sigma^* \right) - \Lambda
&= \frac 2 3 \left( \Delta - N \right) +
\CO\left(\bfrac{m_s}{N_c^2}\right)\quad(205 = 195)
\cr
\Sigma^* - \Sigma &= \Xi^* - \Xi + \CO\left(\bfrac{m_s}{N_c^2}\right)
\quad (191=215)\cr
\frac 3 4 \Lambda + \frac 1 4 \Sigma - \frac 1 2 \left( N + \Xi
\right) &= - \frac 1 4 \left( \Omega - \Xi^* - \Sigma^* + \Delta
\right) + \CO\left(\bfrac{m_s^{3/2}}{N_c^2}\right)\quad(6.5=3.4)
\cr
\frac 1 2 \left( \Sigma^* - \Delta \right) + \frac 1 2 \left( \Omega -
\Xi^* \right)&=\left( \Xi^* -
\Sigma^* \right) +
\CO\left(\bfrac{m_s^{2}}{N_c^2}\right)\quad(145.8=148.8) \cr
\frac 3 4 \Lambda + \frac 1 4 \Sigma &= \frac 1 2 \left( N + \Xi
\right) + \CO\left(\bfrac{m_s^{3/2}}{N_c}\right)\quad(1135=1128.5)\cr
}}
The experimental values for these relations (in MeV) are shown in
parentheses.

The $1/N_c$ expansion helps explain the relative accuracies of the
$SU(3)$ mass relations.  Naively, equal spacing rule II
should work to order $m_s^2$. Since $SU(3)$ mass splittings in the
baryons are of order 150~MeV, and each additional factor of $m_s$ leads
to a suppression by about 25\%, one expects that the equal spacing rule II
is satisfied to about 35~MeV.  Instead, this relation works to about 3~MeV,
since the violation of the equal
spacing rule II is suppressed by an additional factor of $1/N_c^2$, which
makes the relation work ten times better than naively expected. One can
examine the other mass relations in a similar manner, and see that
that the $1/N_c$ expansion explains why some mass relations work
much better than others. There is one relation that does not fit into
this pattern, the Gell-Mann--Okubo formula, which works better than
expected, and implies that $m_2+m_4$ is small \kaplan.
It works as well as equal spacing rule II, which is
violated at $1/N_c^2$, whereas the Gell-Mann--Okubo formula is violated
at order $1/N_c$. The equal spacing rule I also works much better than
expected. However, this can be understood in the $1/N_c$ expansion,
because eq.~\XIIxv\ relates the violation of the Gell-Mann--Okubo formula
to the violation of the equal spacing rule I; if the Gell-Mann--Okubo formula
works
unexpectedly well, so must equal spacing rule I.

A similar analysis can be performed for the masses of baryons containing a
single heavy quark.  The spin-$1/2$ $SU(3)$ flavor ${\bf \bar 3}$ and the
spin-$1/2$ and spin-$3/2$ ${\bf 6}$'s of heavy quark baryons contain eight
isospin multiplets. At order $1/N_c$, these masses are parametrized by seven
operators: $1$, $K$, $K^2$, $I^2$, $J^2$, $J \cdot S_Q$ and $I \cdot S_Q$,
where $S_Q$ is the spin of the heavy quark $Q$.  Thus, there is only a single
mass relation at this order
\eqn\hqsx{
\Sigma_Q^* - \Sigma_Q = \Xi_Q^* - \Xi^\prime_Q  +\CO\left({1\over
N_c^2}\right)\ ,
}
which is the analogue of eq.~\XIIxiv\ for heavy quark baryons.  In
addition, there is a mass relation which relates the heavy quark baryon
hyperfine splittings to the ordinary baryon hyperfine splittings
\callan\ref\jmw{E. Jenkins, A.V. Manohar and M.B. Wise, \np{396}{1993}{27}
\semi Z. Guralnik, M. Luke and A.V. Manohar, \np{390}{1993}{474} \semi
E. Jenkins and A.V. Manohar, \pl{294}{1992}{273}},
\eqn\dn{
\frac 1 3 \left( \Sigma_Q + 2 \Sigma_Q^* \right) - \Lambda_Q
= \frac 2 3 \left( \Delta - N \right)+\CO\left({1\over N_c^2}\right).
}

\newsec{$SU(3)$ Induced Representations}

The previous sections have discussed the meson-baryon couplings using
$SU(2)$ induced representations. This formulation is useful for
understanding the implications of the $1/N_c$ expansion for $SU(3)$
breaking in the baryon sector. In the $SU(3)$ limit, one can combine the
$SU(2)$ analysis with an $SU(3)$ tensor analysis, as was done in Sect.~10,
to determine the meson-baryon couplings in the $SU(3)$ limit. In this
section, we will determine the meson-baryon couplings in the $SU(3)$ limit
by starting directly with an $SU(3)$ invariant formalism for the induced
representations.  Most of the analysis is
identical to the $SU(2)$ analysis of Sect.~4, but there are several
complications which occur in the $SU(3)$ analysis which are absent for $SU(2)$.

Let $\ket{X_0^{iA},\ldots}$ denote states which are the $SU(3)$ analogues of
the
states defined in eq.~\IVi,\footnote{${}^\dagger$}{We will use an uppercase
flavor index $A=1,\ldots,8$ for the $SU(3)$ $X_0$'s, and a lower case flavor
index
$a=1, 2, 3$ for the $SU(2)$ $X_0$'s.}
\eqn\mmi{
\left(X_0^{iA}\right)_{\rm op}  \ket{X_0^{iA},\ldots}
= X_0^{iA}\ket{X_0^{iA},\ldots}.
}
The commutation relations
\eqn\mmii{
\left[J^i, X_0^{jB}\right]=i\,\epsilon_{ijk}\ X_0^{kB},\qquad
\left[T^A,X_0^{jB}\right]=i\,f_{ABC}\ X_0^{jC}
}
imply that $X_0$ transforms as $(1,{\bf 8})$ under $SU(2)_{\rm
spin}\otimes SU(3)_{\rm flavor}$. For a finite spin $\otimes$ flavor
transformation $(g,h)$, $X_0$ transforms as
\eqn\mmiii{\eqalign{
U_J(g)^\dagger\ X_0^{iA}\  U_J(g) &= D_{ij}^{(1)}(g)\  X_0^{jA},\cr
U_T(h)^\dagger\  X_0^{iA}\  U_T(h) &= D_{AB}^{({\bf 8})}(h)\  X_0^{jB},\cr
}}
where $D^{(1)}$ is a representation matrix in the adjoint representation
of $SU(2)$, and $D^{(\bf 8)}$ is a representation matrix in the adjoint
representation of $SU(3)$. The states $\ket{X_0^{iA},\ldots}$ are in one-to-one
correspondence with the space of $3\times 8$ matrices. The irreducible
representations are orbits in the space of matrices under the
transformations eq.~\mmiii. Large $N_c$ baryons correspond to
orbits which contain the element\footnote{${}^\dagger$}{This can be
shown using the methods of Sect.~5.}
\eqn\mmiv{
\bar X_0^{iA} = \cases{1,&$i=A$ and $A\le 3$,\cr 0,&$A>3$.\cr}
}
The little group of $\bar X_0^{iA}$ is $SU(2)\times U(1)\times Z_2$,
where $SU(2)$ is generated by ${\bf K=I+J}$, $U(1)$ is generated by
$T^8$, and $Z_2$ is generated by a $2\pi$ space rotation. The $Z_2$
factor splits the induced representations into fermionic and bosonic
sectors (as for $SU(2)$), and will be omitted from now on. The $\ldots$ in
the state $\ket{\bar X_0^{iA},\ldots}$ is specified by the transformation
properties of $\ket{\bar X_0^{iA},\ldots}$ under the little group
$SU(2)\times U(1)$. The lowest lying baryons containing only $u$, $d$
and $s$ quarks are singlets under the $SU(2)$ group generated by ${\bf
K}$, and have $U(1)$ charge $N_c/\sqrt{12}$. Let us define the states
$\ket{\bar X_0^{iA},y}$ by
\eqn\mmv{\eqalign{
{\bf K}\ \ket{\bar X_0^{iA}, y}&=0,\cr
T^8\  \ket{\bar X_0^{iA}, y}&= N_c\, y\  \ket{\bar X_0^{iA},y},\cr
}}
so that the physical baryons have $y=1/\sqrt{12}$. Then one can define
arbitrary states $\ket{X_0^{iA},y}$ in the $SU(3)$ irreducible representation
by
applying spin and flavor transformations on $\ket{\bar X_0^{iA},y}$. One
can show that all matrices $X_0^{iA}$ on the orbit of $\bar X_0^{iA}$ in
eq.~\mmiv\ can be written as
\eqn\mmvi{
X_0^{iA} = 2\  \Tr A\, T^i\, A^{-1}\, T^A \ ,
}
where $A$ is an $SU(3)$ matrix. Two $A$'s which differ by right
multiplication by a hypercharge transformation, $A\rightarrow A
e^{i\alpha T^8}$ give the same value for
$X_0^{iA}$.\footnote{${}^\ddagger$}{This
discussion closely parallels the quantization of the $SU(3)$ Skyrme
model~\ref\guadagnini{E.~Guadagnini, \np{236} {1984} {35}}.}\ One can define
another operator on the $\ket{X_0^{iA},y}$ basis states which commutes
with $X_0^{iA}$,
\eqn\mmvii{
X_0^{8A} = 2\ \Tr A\, T^8\, A^{-1}\, T^A\ ,
}
which is well defined, and can be written directly in terms of
$X_0^{iA}$ as
\eqn\mmviii{
X_0^{8A} = \bfrac1{\sqrt 3} d_{ABC}\ X_0^{iB}\, X_0^{iC} .
}

The spin and flavor generators on the basis states are
\eqn\mmvix{\eqalign{
&J^i = -i\,\epsilon_{ijk}\ X_0^{jA} {\partial\over\partial X_0^{kA}},\cr
&T^A = -i\,f_{ABC}\ X_0^{iB} {\partial\over\partial X_0^{iC}} + N_c\, y
\ X_0^{8A}.\cr
}}
The flavor generator $T^A$ has two terms, a term which does not commute
with $X_0$ and is of order one, and a term which commutes with $X_0$ and
is of order $N_c$. This feature makes the $N_c$ counting more complicated than
for $SU(2)$, since $T^A$ contains terms which grow with $N_c$.
It is convenient to define
\eqn\mmx{
\hat T^A = -i\,f_{ABC}\ X_0^{iB} {\partial\over\partial X_0^{iC}},
}
so that
\eqn\mmxi{
T^A = \hat T^A + N_c\, y\ X_0^{8A}.
}

The states $\ket{X_0^{iA},y}$ can be decomposed into states with
definite spin and flavor quantum numbers. This analysis is identical to
that in the $SU(3)$ Skyrme model, and will not be repeated here. The
leading term $X_0$ for the meson-baryon couplings gives the axial
current matrix elements (once an overall $N_c$ is factored out) in the
pion sector of order one, in the kaon sector of order $1/\sqrt{N_c}$,
and in the $\eta$ sector of order $1/N_c$. At order $1/N_c$, the
consistency condition eq.~\VIIxi\ has a non-trivial solution for $SU(3)$,
so that to order $1/N_c$,
\eqn\mmxii{
X^{iA} = X_0^{iA} + {\lambda^\prime\over N_c} d_{ABC}\ X_0^{iB}\ T^C.
}
The $d$ symbol vanishes for $SU(2)$, which is why this term was not found
in Sect.~6. The correction in eq.~\mmxii, while formally of order $1/N_c$,
actually makes a contribution of order one, since $T^A$ has a
piece that is of order $N_c$. Using the decomposition eq.~\mmxi\ for
$T^A$, one finds that
\eqn\mmxiii{
X^{iA} = X_0^{iA} + {\lambda^\prime\over N_c} d_{ABC}\ X_0^{iB}
\left(\hat T^C + N_c\, y\ X_0^{8C}\right) \ .
}
The identity
\eqn\mmxiv{
\sqrt 3\ d_{ABC}\ X_0^{iB}\ X_0^{8C} = X_0^{iA}
}
shows that the order one piece is proportional to $X_0$. Thus it is more
convenient to write the $1/N_c$ correction to the axial coupling as
\eqn\mmxv{
X^{iA} = X_0^{iA} + {\lambda\over N_c} d_{ABC}\ X_0^{iB}\ \hat T^C.
}
This correction term yields an order $1/N_c$ correction to the $\CN/\CM$
ratio in eq.~\Xxviii. It produces a correction to the pion couplings at
order $1/N_c^2$, to the kaon couplings at order $1/N_c$ and to the $\eta$
couplings at order one (relative to the leading terms).

\newsec{Chiral Loops}

The $1/N_c$ expansion also allows one to compute corrections to the
$SU(3)$ symmetry limit in a systematic way. The leading corrections are
non-analytic corrections from chiral perturbation
theory.\footnote{${}^\dagger$}{The $\eta^\prime$ mass is of order
$1/\sqrt{N_c}$~\ref\eweta{E.~Witten, \np {156} {1979} {269}\semi G.~Veneziano,
\np {159} {1979} {213}}, and $\eta^\prime$
loops should be included in the large $N_c$ limit. The $\eta^\prime$-nucleon
coupling is of order $1/\sqrt{N_c}$, so $\eta^\prime$ loops contribute at order
$1/N_c$, and are not important for the results discussed in this section.}
\ Naively, these
corrections grow with $N_c$ because the pion-baryon couplings diverge
like $\sqrt{N_c}$. We will see in this section, that the large $N_c$
consistency conditions imply that the corrections to the axial currents
decrease as $1/N_c$, instead of increasing as $N_c$. The $m_s^{3/2}$
corrections to the baryon masses are more interesting.
The order $N_c m_s^{3/2}$ contribution to the baryon masses is
$SU(3)$ singlet, the $m_s^{3/2}$ contribution is $SU(3)$ octet, and it is
only the $m_s^{3/2}/N_c$ contribution that produces a correction to the
$SU(3)$ mass-relations, such as the Gell-Mann--Okubo formula, which are
derived under the assumption of octet symmetry breaking. This pattern confirms
earlier suggestions \kaplan\ref\jaffe{R.L.~Jaffe, \physrev {D21} {1980}
{3215}}\ref\sainio{J.~Gasser, H.~Leutwyler, and M.E.~Sainio, \pl
{253} {1991}{260}}\ref\jmsigma{E.~Jenkins and A.V.~Manohar,
\pl{281}{1992}{336}}\ that the baryon masses might have a strong
non-linear dependence on $m_s$, but that this non-linearity is such that
it does not violate the Gell-Mann--Okubo formula. The chiral corrections
to the baryon magnetic moments are more subtle, and will be discussed
elsewhere.

\subsec{Axial Currents}

The leading non-analytic correction to the baryon axial currents is a
$M^2 \ln M^2$ correction from the loop diagrams shown in \fig\fccc{The
diagrams for the one-loop correction to the baryon axial currents.}. The
renormalization of $X^{iA}$ is proportional to
\eqn\nni{
N_c \left[ X^{jC}, \left[ X^{jB},X^{iA}\right]\right] I^{BC},
}
where the integral $I^{BC}$ depends on the meson masses, and breaks $SU(3)$
symmetry. $I^{BC}$ is equal to $M_\pi^2 \ln M_\pi^2/\mu^2$ for the pions,
$M_K^2 \ln M_K^2/\mu^2$ for the kaons, and $M_\eta^2 \ln
M_\eta^2/\mu^2$ for the $\eta$, and can be written as a linear
combination of $\delta_{BC}$, $d_{BC8}$ and $d_{B88}\, d_{C88}$. Using the
axial
couplings to order $1/N_c$ given by eq.~\mmxv, one finds that
\eqn\nnii{
\left[ X^{jC}, \left[ X^{jB},X^{iA}\right]\right] = \CO\left({1\over
N_c^2}\right),
}
so that the $M^2\ln M^2$ correction is of order $1/N_c$. Thus
meson-loop corrections in the baryon sector are suppressed by $1/N_c$,
just as they are in the meson sector. It is important to keep in mind
that the suppression in eq.~\nnii\ occurs only if one uses
 the large $N_c$ definition of $X$, \ie\ evaluating loop
graphs including the
complete large $N_c$ tower of intermediate states, and using axial couplings
with ratios determined consistently in large $N_c$. This cancellation of the
one-loop chiral logarithmic correction to the axial couplings is precisely what
was found in earlier calculations for $N_c =3$ \ejamaxial.

For $SU(2)$ pion couplings, one has a stronger constraint on the chiral
logarithmic correction than that of eq.~\nnii. Because the order $1/N_c$
correction to $X_0^{ia}$ must be proportional to $X_0^{ia}$, one has the
constraint
\eqn\nnci{
\left[ X^{jb}, \left[ X^{jb},X^{ia}\right]\right] \propto {1\over N_c^2}
X_0^{ia},
}
which fixes the form of the $1/N_c^2$ term in the double commutator.

\subsec{Baryon Masses}

The leading non-analytic correction to the baryon masses is a $M^3$
correction from the graph in \fig\fmmm{The diagram for the $m_s^{3/2}$
correction to the baryon masses.}. The loop graph is proportional
to
\eqn\nniii{
N_c \ X^{iA} X^{iB}\ I^{AB}
}
where $I^{AB}$ is equal to $M_\pi^3$ for the pions, $M_K^3$ for the
kaons, and $M_\eta^3$ for the $\eta$, and can be written as a linear
combination of $\delta_{AB}$, $d_{AB8}$ and $d_{A88}\, d_{B88}$. The
form of the non-analytic correction can then be determined using
eq.~\mmxv\ for $X^{iA}$, and $SU(3)$ identities for the $d$-symbols.

A much simpler way of determining the form of the non-analytic
corrections is to use the $SU(2)$ formalism of Sects.~6--9. The pion
loop corrections are proportional to
\eqn\nniv{
N_c\ g(K)^2\ X^{ia} X^{ia}\ M_\pi^3 = 3 N_c\ g(K)^2\ M_\pi^3
}
using $X^{ia}X^{ia}=3$ (note that $X^{ia}$ is now an $SU(2)$ $X$). Using
eq.~\Xxxv\ for $g(K)$, one finds that the order $N_c$ term from the pion
loops is a constant shift in the baryon mass, and the order one term
(from the $1/N_c$ term in $g(K)$) is a correction proportional to the
number of strange quarks, and so is an $SU(3)$ singlet plus octet.
The  kaon loop corrections are proportional to
\eqn\nnv{
c^2 \left( Y^{i\alpha} Y^{\dagger i\alpha}+ Y^{\dagger i\alpha}
Y^{i\alpha}\right)  M_K^3.
}
The form of $Y$ in eq.~\VIIIxiv\ shows that the order one kaon correction
is of the form of a constant plus a term linear in $K$, and so is a $SU(3)$
singlet plus octet. The $\eta$ loops are of order $1/N_c$. Thus the
order $N_c$ correction to the masses is $SU(3)$ singlet, the order one
piece is singlet plus octet, and the first non-trivial correction first
occurs at order $1/N_c$. A similar result can also be derived for the
non-analytic chiral logarithmic correction to the baryon masses. The
corrections had to have this form because the baryon mass formula eq.~\XIIxii\
was derived in Sect.~10 without assuming $SU(3)$ symmetry, and so it must be
respected by the non-analytic corrections.

\subsec{The Large $N_c$ and Chiral Limits}

Pion loop graphs such as the $M_\pi^3$ contribution to the nucleon mass shown
in \fmmm\ include the entire baryon tower as intermediate states. The baryon
mass splittings $\Delta M$ are of order $1/N_c$. If one first takes the large
$N_c$ limit $N_c\rightarrow\infty$ and then the chiral limit
$m_q\rightarrow 0$, the entire tower of baryons contributes to the non-analytic
$m_q^{3/2}$ mass correction, whereas if one first takes the chiral limit
$m_q\rightarrow 0$ and then the $N_c\rightarrow\infty$ limit, only the nucleon
intermediate state contributes~\ref\jgasser{J.~Gasser, \ap {136} {1981} {62}}.

The non-commutativity of limits does not imply that there is a conflict between
the large $N_c$ and chiral expansions. Chiral perturbation theory is valid
provided $\Delta M$ and $M_\pi$ are small compared with $\Lambda_\chi\approx
1$~GeV, the scale of chiral symmetry breaking, irrespective of the value of
$M_\pi/\Delta M$. The dependence of the nucleon mass on $\Delta M$ and $M_\pi$
is calculable from the graph of \fmmm. The result is of the form
\eqn\zzzxxx{
{1\over 16\pi f^2_\pi} M_\pi^3\ \ F\left({M_\pi\over\Delta M}\right),
}
where $F(x)$ is known~\hungary. The function $F(x)$ has the correct limiting
behavior as $x\rightarrow 0$ and $x\rightarrow\infty$ to correctly reproduce
both the $(N_c\rightarrow\infty, m_q\rightarrow 0)$ and $(m_q\rightarrow 0,
N_c\rightarrow\infty)$ limits. In the real world, $m_q\not=0$ and
$N_c\not=\infty$, and one should evaluate $F(x)$ at the physical value of
$M_\pi/\Delta M$.
\bigskip

The $1/N_c$ expansion gives a systematic method of organizing the chiral
corrections in the baryon sector, so that the non-analytic corrections
are under control. Potentially large corrections either vanish, as
for the axial currents, or can be reabsorbed into lower order terms in
the Lagrangian, as for the masses. It will take a lot more work to see
whether the $1/N_c$ expansion can be combined with
baryon chiral perturbation theory to analyze the
baryons properties in a systematic and controlled expansion.

\newsec{Conclusions}

The $1/N_c$ expansion provides a systematic expansion scheme for
baryons.  The contracted spin-flavor algebra for the baryon sector is
sufficient to constrain the leading and subleading in $1/N_c$ contributions to
various baryon static properties.  The general expansion of operators such as
$X$ and the baryon masses is in powers of $X$, $J/N_c$, $I/N_c$ (or $\hat
T/N_c$ for exact flavor $SU(3)$). The coefficients of the operators have an
expansion in powers of $1/N_c$ and $K/N_c$.
In this work, we have shown that it is possible to consistently extend the
$1/N_c$ expansion to the case of $N_f >2$ light flavors even though the large
$N_c$ flavor representations are different from those for $N_c =3$.  The
extension to more than two light flavors is more involved because of
``representation effects'', which must be taken into account before determining
the predictions for $N_c=3$.

The form of the $1/N_c$ corrections shows that the $N_c\rightarrow\infty$ limit
should be taken with $I$, $J$ and $K$ held fixed. There is no $1/N_c$ expansion
for states with $J$ of order $N_c$. Results for finite $N_c$ are obtained by
expanding about the $N_c\rightarrow\infty$ limit, and using the infinite tower
of baryon states. All effects accounting for the fact that $N_c$ is finite
appear through $1/N_c$ suppressed operators. For instance, for finite $N_c$,
the baryon tower terminates at $J=N_c/2$. The finite height of the tower away
from $N_c\rightarrow\infty$ results in $1/N_c$ corrections to calculations
performed with the infinite tower. These corrections are automatically included
in the $1/N_c$ suppressed operators.

Many of the results obtained in the $1/N_c$ expansion for baryons are the same
as those obtained in the Skyrme or non-relativistic quark models. The results
obtained using the $1/N_c$ expansion are those model relations that work
``well,'' such as $F/D$ ratios. The operator structure of the $1/N_c$ expansion
is similar to that in the Skyrme and quark models. In large $N_c$ QCD, the
coefficients of the different operators are not determined by the consistency
conditions. The models, on the other hand, make definite predictions for these
coefficients. Model dependent relations which depend on the values of these
coefficients, such as the absolute normalization of $g_A$, do not work well,
and can not be derived from large $N_c$ QCD. The non-relativistic quark model
has a $SU(2N_f)$ spin-flavor symmetry for mesons as well as baryons. Only a
contracted $SU(2N_f)$ spin-flavor symmetry for baryons exists in the $1/N_c$
expansion.

The $1/N_c$ expansion also provides a way of computing chiral loops in
the baryon  sector. Earlier computations of chiral loops were plagued by
large non-analytic corrections \ref\gasser{J.~Gasser and H.~Leutwyler,
\ap{158} {1984} {142}, \np{250} {1985} {465}}\ref\wise{J.~Bijnens,
H.~Sonoda and M.B.~Wise, \np{261} {1985} {185}}. The $1/N_c$ expansion
provides an alternative calculational scheme in which the
entire degenerate baryon tower is included as intermediate states in loop
diagrams. This procedure
leads to large cancellations, and makes the chiral expansion better behaved.
Large corrections which do not cancel can be reabsorbed into lower order
parameters.

Whether the $1/N_c$ expansion proves useful depends on the size of the $1/N_c$
corrections. The corrections appear to be under control for the baryon axial
currents and masses.  In the meson sector, there is one example where the
$1/N_c$ corrections are large, the
$\Delta I=1/2$ rule \ref\deltahalf{M.~Fukugita,
T.~Inami, N.~Sakai, and S.~Yazaki, \pl{72} {1977} {237}\semi
R.S.~Chivukula, J.M.~Flynn, and
H.~Georgi, \pl{171} {1986} {453}}. In the large $N_c$ limit,
factorization is exact, and the $\Delta I=1/2$ and $\Delta I=3/2$ amplitudes
are related by a Clebsch-Gordan coefficient, with no large
enhancement of the $\Delta I=1/2$ amplitude. However, other large $N_c$
predictions such as Zweig's rule hold in the meson sector.

\centerline{{\bf Acknowledgements}}

We wish to thank J.~Bijnens, H.~Georgi, Z.~Guralnik, D.B.~Kaplan, J.~Kuti,
M.~Saadi, M.B.~Wise and K.~Yamawaki for discussions.
We have recently received papers on large $N_c$ baryons by C.~Carone, H.~Georgi
and S.~Osofsky \ref\carone{C.~Carone, H.~Georgi
and S.~Osofsky, HUTP-93/A032, {\tt [hep-ph/9310365]}}, and by M.A.~Luty and
J.~March-Russell \ref\marchrussell{M.A.~Luty and J.~March-Russell, LBL-34778,
{\tt [hep-ph/9310369]}}.
This work was supported in part by the Department of Energy under grant number
DOE-FG03-90ER40546 and the Texas National Research Laboratory Commission under
grant RGFY93-206.  A.M. was also supported by PYI award PHY-8958081.
E.J. and A.M. thank the Aspen Center for Physics for hospitality while
some of this work was completed.

\listrefs
\listfigs
\insertfig{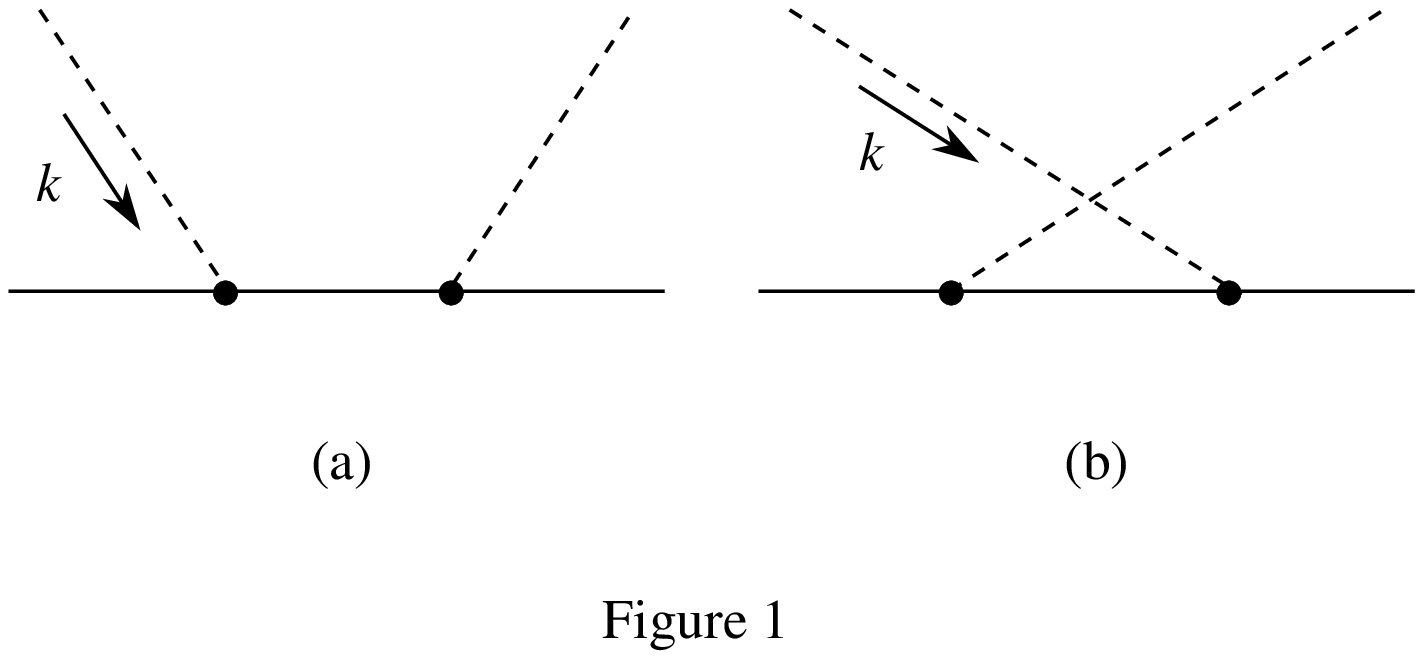}
\insertfig{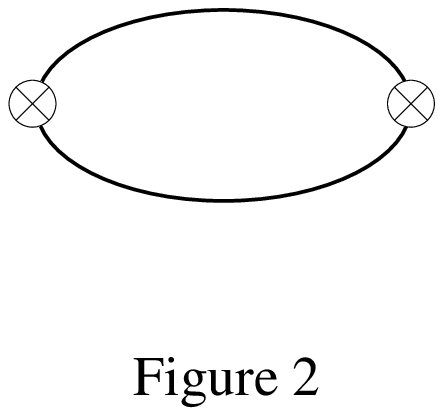}
\insertfig{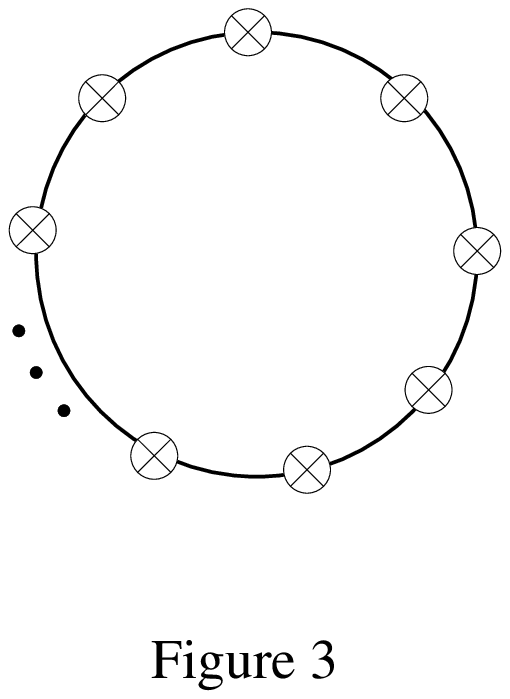}
\insertfig{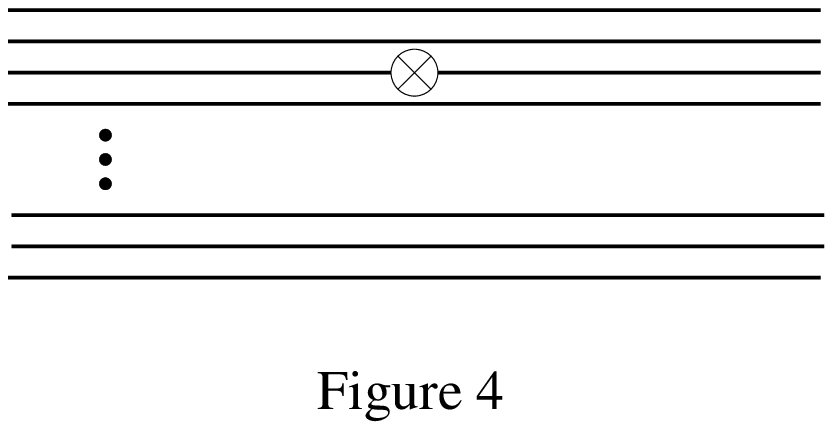}
\insertfig{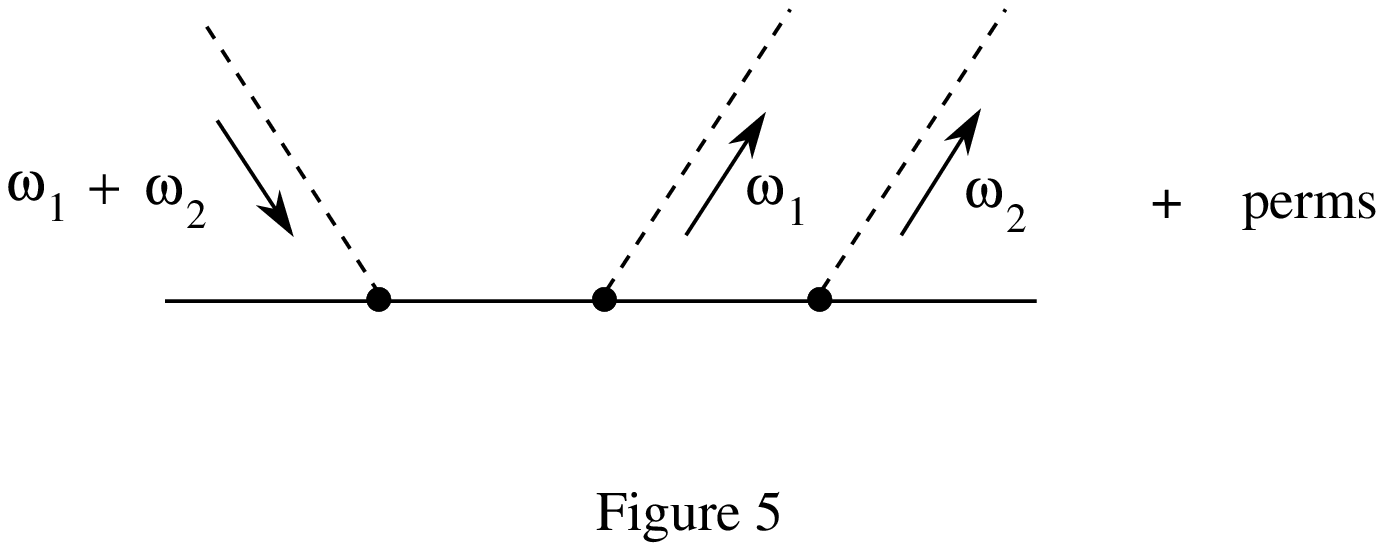}
\insertfig{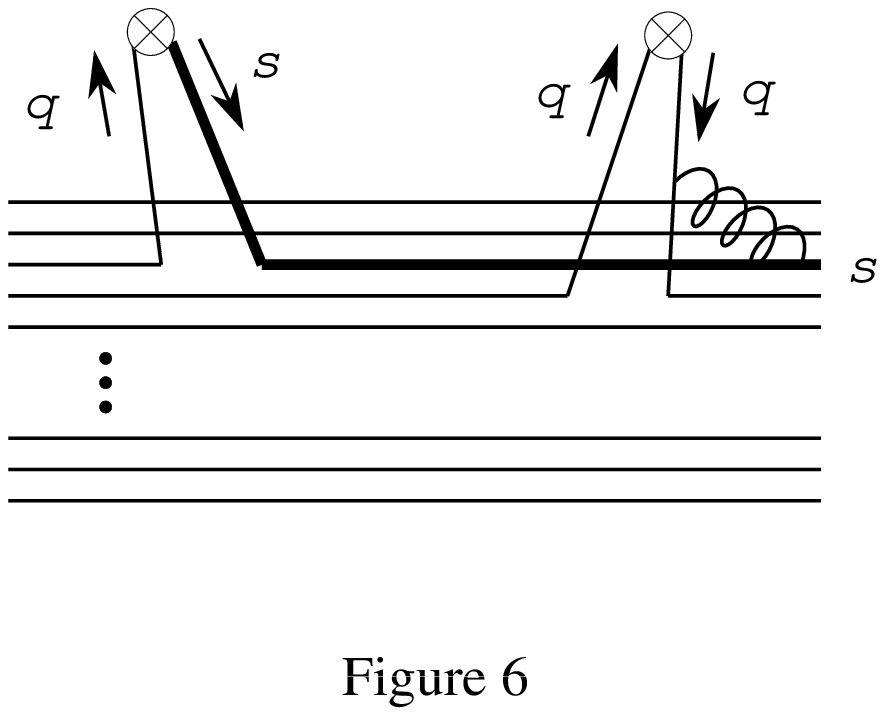}
\insertfig{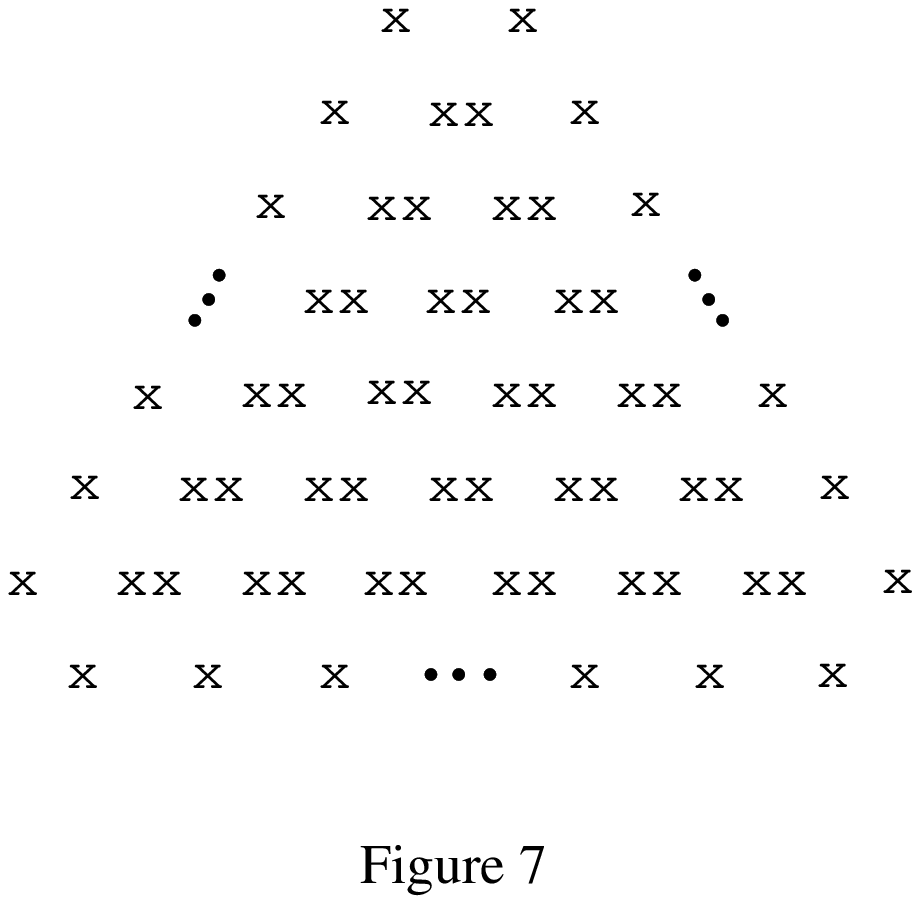}
\insertfig{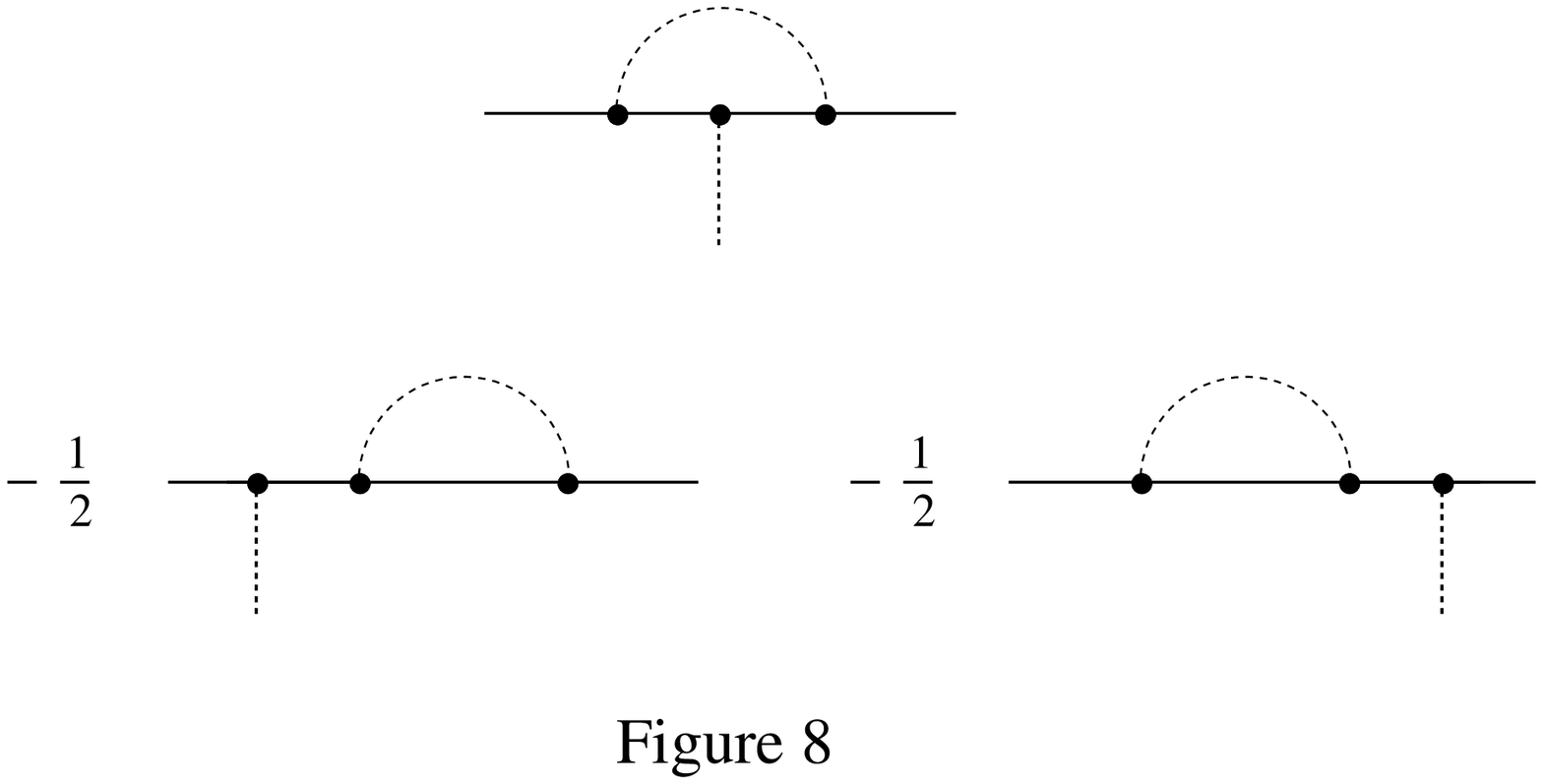}
\insertfig{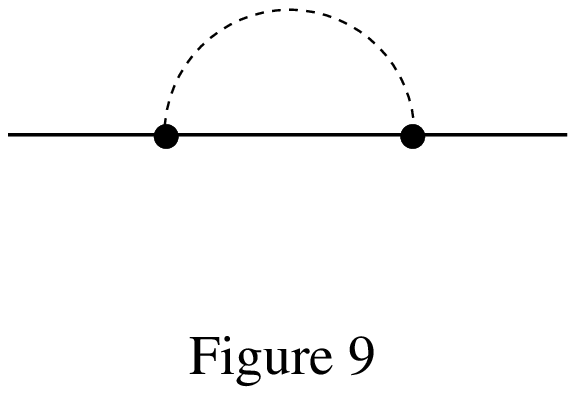}

\bye